\documentclass[review,3p]{elsarticle}

\usepackage{blindtext}
\usepackage[T1]{fontenc}
\usepackage[utf8]{inputenc}
\usepackage[font=small]{caption}
\usepackage{verbatim}
\usepackage{float}
\usepackage{amsmath}
\usepackage{amsthm}
\usepackage{amssymb}
\usepackage{enumitem}
\usepackage{siunitx}
\usepackage{mathdots}
\usepackage{graphicx}
\usepackage{fancyhdr}
\usepackage{amsfonts}
\usepackage{fixmath}
\usepackage{nicefrac}
\usepackage{resizegather}
\usepackage{ulem}
\usepackage{wasysym}
\usepackage{xcolor}

\usepackage{mdframed}
\newmdenv[
topline=false,
leftline=false,
rightline=false,
bottomline=false,
]{txtbox}

\usepackage{caption}
\usepackage{ragged2e}
\usepackage{booktabs}
\usepackage{array}

\usepackage{hyperref}
\hypersetup{colorlinks,citecolor=black,urlcolor=black,linkcolor=black} 
\newcommand{\ie}{\textit{i.e.}}
\newcommand{\eg}{\textit{e.g.}}

\newcommand{\dbscan}{DBSCAN-WGL}
\newcommand{\undbscan}{DBSCAN-Unweighted}
\newcommand{\cdp}{CDP-WGL}
\newcommand{\uncdp}{CDP-Unweighted}
\newcommand{\meanshift}{MeanShift-WGL}
\newcommand{\unmeanshift}{MeanShift-Unweighted}

\usepackage{multirow}

\usepackage{algorithm}
\usepackage[beginComment=//~, beginLComment=//~, endLComment=, commentColor=black, italicComments=true, indLines=false, noEnd=true]{algpseudocodex}
\algnewcommand\algorithmicforeach{\textbf{for each}}
\algdef{S}[FOR]{ForEach}[1]{\algorithmicforeach\ #1\ \algorithmicdo}
\renewcommand{\LComment}[1]{\Statex \textit{// #1}}

\usepackage{etoolbox}
\usepackage{tikz}
\usetikzlibrary{tikzmark}
\usetikzlibrary{calc}

\errorcontextlines\maxdimen
\newcommand{\ALGtikzmarkcolor}{black}
\newcommand{\ALGtikzmarkextraindent}{4pt}
\newcommand{\ALGtikzmarkverticaloffsetstart}{-.5ex}
\newcommand{\ALGtikzmarkverticaloffsetend}{-.5ex}
\makeatletter
\newcounter{ALG@tikzmark@tempcnta}

\newcommand\ALG@tikzmark@start{%
    \global\let\ALG@tikzmark@last\ALG@tikzmark@starttext%
    \expandafter\edef\csname ALG@tikzmark@\theALG@nested\endcsname{\theALG@tikzmark@tempcnta}%
    \tikzmark{ALG@tikzmark@start@\csname ALG@tikzmark@\theALG@nested\endcsname}%
    \addtocounter{ALG@tikzmark@tempcnta}{1}%
}

\def\ALG@tikzmark@starttext{start}
\newcommand\ALG@tikzmark@end{%
    \ifx\ALG@tikzmark@last\ALG@tikzmark@starttext
    \else
        \tikzmark{ALG@tikzmark@end@\csname ALG@tikzmark@\theALG@nested\endcsname}%
        \tikz[overlay,remember picture] \draw[\ALGtikzmarkcolor] let \p{S}=($(pic cs:ALG@tikzmark@start@\csname ALG@tikzmark@\theALG@nested\endcsname)+(\ALGtikzmarkextraindent,\ALGtikzmarkverticaloffsetstart)$), \p{E}=($(pic cs:ALG@tikzmark@end@\csname ALG@tikzmark@\theALG@nested\endcsname)+(\ALGtikzmarkextraindent,\ALGtikzmarkverticaloffsetend)$) in (\x{S},\y{S})--(\x{S},\y{E});%
    \fi
    \gdef\ALG@tikzmark@last{end}%
}

\apptocmd{\ALG@beginblock}{\ALG@tikzmark@start}{}{\errmessage{failed to patch}}
\pretocmd{\ALG@endblock}{\ALG@tikzmark@end}{}{\errmessage{failed to patch}}
\makeatother

\usepackage{resizegather}

\usepackage{xcolor}

\theoremstyle{definition}
\newtheorem{remark}{Remark}

\newtheorem{example}{Example}

\makeatother

\journal{Signal Processing}
\begin{document}

\begin{frontmatter}

\title{Distributed Multi-Object Tracking Under Limited Field of View Heterogeneous Sensors with Density Clustering}

\author[1]{Fei Chen\corref{cor2}}
\ead{fei.chen@adelaide.edu.au}
\cortext[cor2]{First co-authors in alphabetical order and contributed equally to this work}

\author[2]{Hoa Van Nguyen\corref{cor2}}
\ead{hoa.v.nguyen@curtin.edu.au}

\author[3]{Alex S. Leong}
\ead{alex.leong@defence.gov.au}

\author[3]{Sabita Panicker}
\ead{sabita.panicker1@defence.gov.au}

\author[4]{Robin Baker}
\ead{robin.baker@defence.gov.au}

\author[1]{Damith C. Ranasinghe\corref{cor1}}
\ead{damith.ranasinghe@adelaide.edu.au}
\cortext[cor1]{Corresponding author}

\address[1]{School of Computer and Mathematical Science, The University of Adelaide, Adelaide, SA 5005, Australia}
\address[2]{School of Electrical Engineering, Computing, and Mathematical Sciences,
Curtin University, Bentley, WA 6102, Australia}
\address[3]{Platforms Division, Defence Science and Technology Group, Australia}
\address[4]{Land and Integrated Force Division,  Defence Science and Technology Group,  Edinburgh, SA, Australia}

\begin{abstract}
We consider the problem of tracking multiple, unknown, and time-varying numbers of objects using a distributed network of heterogeneous sensors. In an effort to derive a formulation for practical settings, we consider \textit{limited} and \textit{unknown} sensor field-of-views (FoVs), sensors with limited local \textit{computational resources} and \textit{communication channel capacity}. The resulting distributed multi-object tracking algorithm involves solving an NP-hard multidimensional assignment problem either optimally for small-size problems or sub-optimally for general practical problems. For general problems, we propose an efficient distributed multi-object tracking algorithm that performs track-to-track fusion using a clustering-based analysis of the state space transformed into a density space to mitigate the complexity of the assignment problem. The proposed algorithm can more efficiently group local track estimates for fusion than existing approaches. To ensure we achieve globally consistent identities for tracks across a network of nodes as objects move between FoVs, we develop a graph-based algorithm to achieve label consensus and minimise track segmentation. Numerical experiments with synthetic and real-world trajectory datasets demonstrate that our proposed method is significantly more computationally efficient than state-of-the-art solutions, achieving similar tracking accuracy and bandwidth requirements but with improved label consistency.
\end{abstract}

\end{frontmatter}

\section{Introduction}\label{sec:intro}
Multi-Object Tracking (MOT) aims to detect, classify, and estimate the trajectories of an unknown and time-varying number of objects using noisy sensor measurements. The objects of interest can be either stationary, such as mines~\cite{williams2010optimal}, mobile, such as vehicles~\cite{braca2015distributed}, or hybrid, such as offshore critical infrastructures~\cite{caiti2013mobile}. In particular, MOT involves estimating the trajectories of the objects and maintaining their provisional identities or labels. Trajectories are important for capturing the behaviour of the objects, while labels provide the means to distinguish individual object trajectories and allow human/machine users to communicate information regarding the relevant objects of interest, consistently. Solving MOT problems is challenging due to noisy measurements, false alarms, misdetections, and unknown data association (i.e., unknown measurement-to-object association). Despite these challenges, MOT plays significant roles in various applications and domains such as surveillance~\cite{blackman1999design,bar2011tracking}, aerospace~\cite{reid1979algorithm}, cell biology~\cite{rezatofighi2015multi,dat2021celltracking}, robotics~\cite{hoa2019jofr,hoa2020iros,fei2023jofr}, and computer vision~\cite{cox1996an,tian2024,linh2024}. 

General MOT algorithms can be categorised into three main frameworks: Multiple Hypothesis Tracking (MHT)~\cite{blackman1999design,bar2011tracking,reid1979algorithm}, Joint Probabilistic Data Association (JPDA)~\cite{rezatofighi2015joint,bar2011tracking,musicki2004joint}, and Random Finite Sets (RFS)~\cite{mahler2007statistical,mahler2014advances}. Traditional approaches like MHT and JPDA solve the measurement-to-object assignment problem before estimating object states using a single-object filter. JPDA computes association probabilities between objects and measurements by considering their statistical properties~\cite{fortmann1983sonar,chang1986joint}. However, JPDA lacks a mechanism for handling object birth and death, requiring separate procedures for managing a time-varying number of objects~\cite{rezatofighi2015multi,svensson2011set}. MHT addresses data association challenges by forming and evaluating multiple hypotheses on object-to-measurement associations over multiple scans or frames~\cite{blackman1999design,reid1979algorithm}, with notable approaches like Hypothesis-Oriented MHT (HOMHT)~\cite{danchick1993fast,cox1995finding} and Track-Oriented MHT (TOMHT)~\cite{bar2007dimensionless,coraluppi2014if,chong2019forty}. RFS, a newer approach, directly models the multi-object state as a set-valued random variable, enabling estimation without explicitly solving the data association problem~\cite{mahler2007statistical} such as the Probability Hypothesis Density (PHD)~\cite{mahler2003multitarget} and Cardinalized PHD (CPHD)~\cite{mahler2007cphd} filters, the Multi-Bernoulli (MB) filter~\cite{mahler2007statistical,vo2008cardinality}, the Poisson Multi-Bernoulli Mixture (PMBM) filter~\cite{williams2015marginal}, the Labelled Multi-Bernoulli (LMB) filter~\cite{reuter2014lmb}, and the Generalised Labelled Multi-Bernoulli (GLMB) filter~\cite{vo2013glmb,vo2014glmb}---the first exact closed-form multi-object tracking algorithm.

Recently, Wireless Sensor Networks (WSN) %such as Anti-Submarine Warfare (ASW) 
systems with interconnected nodes of underwater sonobuoys~\cite{georgy2011clustered}, and networked Autonomous Underwater Vehicles (AUV)~\cite{ferri2017cooperative} equipped with sensing, communication, and processing capabilities, have garnered significant research interest for MOT applications. These networks address the practical challenge of limited field of view (FoV) from a single node, crucial in scenarios with objects dispersed over large areas~\cite{hu2020decentralized}, such as tracking air traffic~\cite{wang1999}, visual tracking of wildlife~\cite{betke2007}, and monitoring space debris~\cite{jones2015, klinkrad2006, nrc2012}. 
Multiple sensors in a network facilitate more accurate tracking by fusing the common information and generating holistic multi-object trajectories by augmenting exclusive information gathered by FoV-limited sensors at each node. 

The benefits of sensor networks for MOT motivate the investigation of distributed MOT (DMOT). Unlike centralised MOT, where measurements are sent to a fusion centre, DMOT allows each sensor node to operate independently, offering scalability, flexibility, and robustness to node failures~\cite{luo2006dist,li2017flooding}.
However, DMOT is a non-trivial problem~\cite{uney2013dist}. It encompasses the challenges inherent to an MOT problem, such as the appearance and disappearance of objects, false measurements,  misdetections, and uncertainties in measurement-to-object association~\cite{mahler2007statistical,vo2013glmb}. Further, DMOT requires solving a complex fusion problem for achieving kinematic and label consensus in a distributed environment where sensors operate independently, possess heterogeneous sensing capabilities~\cite{ferri2017cooperative,hu2020decentralized}, and are limited by computational capabilities and communication bandwidth~\cite{stojanovic2007relationship}.
Information fusion in DMOT is prone to the ``\textit{double counting}'' problem~\cite{fantacci2018robust}, necessitating a suitable fusion algorithm to improve tracking accuracy by fusing common information  while preserving the complementary information from  limited FoV sensors~\cite{hoa2021distributed}.

Recent algorithms, based on the RFS framework, offer a principled generalisation from single object tracking to MOT and DMOT.
RFS-based algorithms for multi-object density fusion
generally follow two main approaches: 
\textit{i)~Geometric Averaging (GA)} (also known as Generalised Covariance Intersection or \textit{GCI}), such as GA-PHD~\cite{uney2013dist,li2020Guchong}, GA-CPHD~\cite{battistelli2013consensus}, GA-MB~\cite{wang2016distributed}, and GA-GLMB/LMB~\cite{fantacci2018robust,li2017robust,li2018computationally}; 
\textit{ii)~Arithmetic Averaging (AA)}, including AA-PHD~\cite{li2017clustering,li2019local,Li2017dphd,LiTian2019partial}, AA-CPHD~\cite{Gao2020multiobject}, AA-MB~\cite{LiTian2020on}, AA-LMB/GLMB~\cite{gao2020fusion}, and AA-PMBM~\cite{li2023best}. 
A recent study indicated that both GA and AA fusion methodologies fundamentally resemble the Fr\'{e}chet means~\cite{li2022multisensor}, defining the central tendency of distributions within arbitrary metric spaces.

The multi-object density fusion techniques offer elegant conceptual solutions but their realisation demands substantial computational and communication bandwidth due to the large number of parameters resulting in a multi-object density,
making them challenging to employ for practical, real-time applications. These challenges become especially pronounced as the number of interconnected nodes and objects continues to increase~\cite{hoa2021distributed,gao2019event}. 
Further, both GA and AA fusion methods experience significant performance degradation when sensors have \textit{disparate FoVs}. This issue is particularly acute for GA, which effectively preserves only objects within the intersection of all sensors' FoVs \cite{Gao2020multiobject,uney2019fusion}. While AA fusion partially mitigates this problem because of its additive nature, it still faces challenges when FoV inconsistencies are not adequately considered~\cite{yi2020distributed}. Although density-based fusion methods with limited FoVs were investigated using GA-PHD in~\cite{li2020Guchong} and AA-PHD in~\cite{li2019local}, the algorithms require \textit{knowing FoV information} of all sensor nodes. A robust AA-PHD fusion method for unknown FoVs was subsequently proposed in \cite{yi2020distributed}. Notably, these aforementioned density fusion methods are mainly employed for \textit{unlabelled} multi-object densities to tackle only \textit{multi-object state estimation problems}, i.e., without trajectory (or label) information.

In the context of DMOT, dealing with \textit{labelled} multi-object density fusion for tracking with networks of sensor nodes, regardless of whether nodes share the same FoV, comparing labelled densities from different nodes is challenging. In particular, labels are discrete random variables defined by users or nodes to distinguish multiple distinct objects and usually vary across different nodes. Standard divergences (e.g., Kullback–Leibler divergence) are undefined for labelled densities with different supports.

An alternative to density fusion is track-to-track fusion. This leads to solving a multi-dimensional assignment problem for multiple nodes. Unfortunately, optimal solutions to the problem have an NP-hard complexity. Although this can be addressed with heuristics or techniques such as sequential pairwise matching~\cite{kaplan2008assignment}, traditional track-to-track fusion assumes zero false tracks and missed tracks, in other words, the number of local tracks from any
 two nodes is assumed to be the same~\cite{chong1990distributed,chang1997on,chong2000architectures,mori2002track,mori2003track,mori2014performance,tian2015track}. However, this is not a realistic assumption in settings with limited FoV sensors and/or when the number of objects is time-varying~\cite{hoa2021distributed}. Importantly, under false and missed tracks, prominent in limited or unknown FoV settings, the label consistency problem is exacerbated as a single label must be reached for fused tracks.

Track-to-track fusion under limited, unknown FoV sensors has recently been investigated in~\cite{hoa2021distributed,he2018multi}. The DMOT method in~\cite{hoa2021distributed} employs pairwise matching to manage the complexity of the multi-dimensional assignment problem and addresses the label consistency problem---ensuring globally consistent labels for the consensus tracks~\cite{hoa2021distributed, li2017clustering, dunham2004multiple}, i.e. achieving globally consistent trajectories at each node to construct a single integrated air picture (SIAP) across a sensor network. In contrast, in~\cite{he2018multi}, complexity is mitigated by using clustering with Density-Based Spatial Clustering of Applications with Noise (DBSCAN)~\cite{ester1996density} algorithm to analyse and fuse shared JPDA filter object densities. Using cluster analysis as a method for mitigating complexity, significantly reduces the computational cost at a node compared to pairwise matching in~\cite{hoa2021distributed}. However, analysis with DBSCAN~\cite{ester1996density} necessitates a prior density threshold---\textit{hyper parameter}---to distinguish between data points in a clustering space, but determining a suitable density threshold is \textit{tracking-scenario dependent} and a non-trivial task~\cite{cheng1995mean}. Importantly, the reliance on densities leads to increasing demands on communication bandwidth, while the label consistency problem in DMOT remains unresolved.

In this study, we propose a new track-to-track fusion method for limited, heterogeneous, unknown FoV sensors to advance DMOT. We mitigate the labelled density fusion problem by considering fusing labelled estimates of local tracks shared by nodes instead of labelled densities. But first, we fuse estimates to achieve kinematic consensus and then tackle the label consistency problem to achieve globally consistent labels for tracks across a sensor network. In achieving kinematic consensus, we alleviate the complexity of the assignment problem that arises in fusing estimates by introducing a contemporary cluster analysis-based approach to overcome the difficulty in identifying suitable density thresholds to efficiently group estimates for fusion. To address the label consistency problem, we introduce a Weighted Graph Label (WGL) algorithm designed to maintain a record of association history between track labels from different nodes to ultimately facilitate label consensus and SIAP. In particular, the approach offers enhanced robustness against outlier labels compared to the prior state-of-the-art method. Importantly, our approach advances DMOT by addressing key challenges in practical problems:
\begin{itemize}
\item \textbf{Sensor heterogeneity and Unknown FoVs:} We fuse labelled estimates instead of labelled multi-object densities. Therefore: i)~our approach is versatile in handling sensor heterogeneity and unknown FoVs, ensuring applicability across diverse practical scenarios; and ii)~importantly, is \textit{agnostic} to the choice of the local MOT filter at each sensor node.
\item \textbf{Limited computational capabilities:} To mitigate the problems faced in fusing labelled multi-object densities (tracks) %, and the inherent complexity of fusing labelled densities, 
we fuse estimates and subsequently solve the resulting label consistency problem. We reduce the computational complexity of the resulting fusion problem by using a more generalisable cluster analysis formulation. Notably, the method we introduce eliminates the difficult problem of identifying suitable density threshold hyper-parameters.
\item \textbf{Bandwidth-limited communication channels:} By fusing labelled estimates instead of densities, we significantly reduce the volume of data exchanged and consequently, demands on communication bandwidth. This makes our technique suitable for practical, bandwidth-limited applications.
\end{itemize}

The remainder of this paper is organised as follows. 
We provide background to aid with the problem formulation and our evaluations in Section \ref{sec:background}.
Section \ref{sec:proposed_method} defines the problem and presents our proposed fusion method and label consensus algorithm.
Section \ref{sec:experiment} details numerical experiments, results and comparisons with existing methods.
Section \ref{sec:conclusion} discusses concluding remarks.

\section{Background}\label{sec:background}
This section provides our notation conventions, the necessary background on some fundamental concepts of MOT within the RFS framework in Section~\ref{sec:lmb}, and background on clustering algorithms in Section~\ref{sec:clustering_algo}.

\subsection{Notation} We follow the notation conventions in \cite{reuter2014lmb}. Lowercase letters (\eg, $x, \mathbf{x}$) denote single-object states, while uppercase letters (\eg, $X, \mathbf{X}$) denote multi-object states. Unbold letters denote unlabelled states and their densities ($X,\pi$), bold letters represent labelled states and their densities ($\mathbf{X}$, $\boldsymbol{\pi}$), while spaces are in blackboard letters ($\mathbb{X}$, $\mathbb{L}$). For a set $X$, $\mathcal{F}(X)$ denotes the class of finite subsets of $X$, $\mathcal{F}_{n}(X)$ denotes the class of finite subsets of $X$ with cardinality $n$, and $1_{X}(\cdot)$ is the indicator function of $X$ whose cardinality is $|X|$. The multi-object exponential $f^X$ for a function $f$ is defined as $\prod_{x \in X} f(x)$, with $f^{\emptyset} = 1$.  We denote the generalised Kronecker delta function by $\delta_Y(X)$, equalling $1$ if $X=Y$ and $0$ otherwise. The inner product $\int f(x) g(x) dx$ is denoted as $\langle f,g \rangle$ for brevity.

We summarise a description of notations in Table~\ref{tab:notations} and abbreviations in Table~\ref{tab:abbrev}.

\begin{table}[!tb]
\centering
\small
\caption{Basic Notations}
\label{tab:notations}
\begin{tabular}{ll}
\toprule
\textbf{Symbol}      & \textbf{Description} \\ 
\hline
$\mathcal{N}$ & set of sensor nodes \\
$G$ & weighted label graph\\
$D$ & clustering index\\
$\mathbb{X}$ & state space \\
$\mathbb{L}$ & label space \\
$\mathbb{I} = \mathbb{L}\times\mathcal{N}$ & global label space \\
$\mathcal{L}: \mathbb{X}\times\mathbb{I}\rightarrow\mathbb{I}$ & label extraction function \\
$\mathcal{S}: \mathbb{X}\times\mathbb{I}\rightarrow\mathbb{X}$ & state extraction function \\
$\mathbf{x} = (x, \ell)$ & labelled single-object state \\
$\mathbf{X}\subset\mathbb{X}\times\mathbb{L}$ & labelled multi-object state \\
$\mathbf{X}^{(n)}\subset\mathbb{X}\times\mathbb{I}$ & set of non-fused local labelled state estimates of node $n$ \\
$M^{(n)}$ & number of non-fused local state estimates of node $n$, i.e. cardinality of $\mathbf{X}^{(n)}$ \\
$\mathbf{X}^{(\text{local})}\subset\mathbb{X}\times\mathbb{I}$ & set of non-fused local labelled state estimates\\
$\mathbf{X}^{(\text{local})}_{m}\subset\mathbb{X}\times\mathbb{I}$ & set of non-fused local labelled state estimates with the same clustering index $m$\\
$\mathbf{X}^{(\text{global})}\subset\mathbb{X}\times\mathbb{I}$ & set of consensed global labelled state estimates \\
$\mathbf{L}^{(\text{local})}_{m}$ & set of non-fused local labels with the same clustering index $m$\\
$\mathbf{L}^{(\text{global})}$ & set of consensed global labels \\
\bottomrule
\end{tabular}
\end{table}

\begin{table}[!tb]
\centering
\caption{Abbreviations}
\label{tab:abbrev}
\begin{tabular}{llll}
\toprule
\textbf{Acronym} & \textbf{Full Texts} & \textbf{Acronym} & \textbf{Full Texts} \\
\midrule
AA & Arithmetic Averaging & MHT & Multiple Hypothesis Tracking \\
AUV & Autonomous Underwater Vehicles & MOT & Multi-Object Tracking \\
CDP & Clustering Density Peak & MS & Multi-Sensor \\
CPHD & Cardinalized Probability Hypothesis Density & OSPA & Optimal Sub-Pattern Assignment \\
DBSCAN & Density-Based Spatial Clustering of & OSPA\textsuperscript{(2)} & OSPA-on-OSPA \\
 & Applications with Noise & PHD & Probability Hypothesis Density \\    
FoV & Field of View & PMBM & Poisson Multi-Bernoulli Mixture \\
GA & Geometric Averaging & RFS & Random Finite Set \\
GLMB & Generalised Labelled Multi-Bernoulli & SIAP & Single Integrated Air Picture \\
JPDA & Joint Probabilistic Data Association & TC & Track Consensus \\
LMB & Labelled Multi-Bernoulli & WGL & Weighted Graph Label \\
MB & Multi-Bernoulli & WSN & Wireless Sensor Networks \\
\bottomrule
\end{tabular}
\end{table}

\subsection{Labelled Multi-Bernoulli (LMB) Filter}\label{sec:lmb}
Solving MOT problems is challenging due to uncertainty in physical sensors, such as noisy measurements, misdetections, false alarms and unknown data association. Despite these formidable challenges, several MOT algorithms have been developed, which can be categorised into three main frameworks: MHT, JPDA and RFS. The RFS framework has gained more recent popularity due to its rigorous mathematical foundation and provides an effective method for managing complex tracking scenarios. This is achieved by treating the state of multiple objects as a finite set and subsequently utilising Finite Set Statistics (FISST) techniques to estimate the evolving set over time. Due to its rigorous mathematical basis, several RFS-based filters have been developed, including the Probability Hypothesis Density (PHD) filter~\cite{mahler2003multitarget}, the Cardinalized Probability Hypothesis Density (CPHD) filter ~\cite{mahler2007cphd}, the Multi-Bernoulli (MB) filter~\cite{mahler2007statistical,vo2008cardinality}, the Poisson Multi-Bernoulli Mixture (PMBM) filter~\cite{williams2015marginal}, the Labelled Multi-Bernoulli (LMB) filter~\cite{reuter2014lmb}, and the Generalised Labelled Multi-Bernoulli (GLMB) filter~\cite{vo2013glmb,vo2014glmb}---the first exact closed-form multi-object tracking algorithm.

In particular, the LMB filter~\cite{reuter2014lmb} serves as an approximation to the GLMB filter, significantly reducing the number of association hypotheses while maintaining reasonable tracking performance. Given our primary focus on developing an efficient DMOT algorithm within the constraints of limited computational resources, we have chosen to employ the LMB filter as our tracking algorithm at local sensor nodes. However, it is important to highlight that our proposed DMOT algorithm is \textit{agnostic to the choice of local MOT filters}, including but not limited to MHT and JPDA. The following presents the necessary background on the LMB filter employed in our evaluations.

\vspace{2mm}
\noindent\textbf{Labelled Multi-Bernoulli (LMB) RFS.~} An LMB RFS $\mathbf{X}$ is fully characterised  by the parameter set $\boldsymbol{\pi} = \{r(\ell),p(\cdot,\ell)\}_{\ell \in \mathbb{L}} $, where  $r(\ell)$ is the label existence probability and $p(\cdot,\ell)$ is the spatial label density, with $\int p(x,\ell)dx = 1$. The LMB density is given by~\cite{reuter2014lmb}
\begin{align} \label{eq_LMB_RFS}
	\boldsymbol{\pi}(\mathbf{X}) = \triangle(\mathbf{X}) w(\mathcal{L}(\mathbf{X}))p^{\mathbf{X}}.
\end{align}

Here, $\mathcal{L}(\mathbf{X}) = \{ \mathcal{L}(\mathbf{x}): \mathbf{x} \in \mathbf{X}\}$  is the set of labels of the labelled RFS $\mathbf{X}$ where $\mathcal{L}: \mathbb{X} \times \mathbb{L} \rightarrow \mathbb{L}$ is the label projection given by $\mathcal{L}(x,\ell) = \ell$, $\triangle(\mathbf{X}) = \delta_{|\mathbf{X}|}(|\mathcal{L}(\mathbf{X})|)$ is an indicator for distinct labels, 
$w(L) = r^{L} (1-r)^{\mathbb{L}\setminus L}$, and $p(\mathbf{x}) = p(x,\ell)$. For brevity, we represent the LMB density as $\boldsymbol{\pi} =\{r(\ell),p(\cdot,\ell)\}_{\ell \in \mathbb{L}} = \big\{ (w(I),p): I \in \mathcal{F}(\mathbb{L}) \big\}$.

\vspace{2mm}
\noindent\textbf{RFS based Multi-object Filtering Theory.}
Given a labelled RFS $\mathbf{X}_{k}$ at time $k$, the corresponding multi-object density can be propagated using prediction and update steps of the Bayes multi-object filter:

\begin{align}
\boldsymbol{\pi}_{k+1|k}(\mathbf{X}_{k+1}|Z_{1:k}) =  \int \Phi_{k+1|k}(\mathbf{X}_{k+1}|\mathbf{X})\boldsymbol{\pi}_{k}(\mathbf{X}|Z_{1:k})\delta\mathbf{X} \label{eq:FISST_predit} \\
\boldsymbol{\pi}_{k+1}(\mathbf{X}_{k+1}|Z_{1:k+1}) = \frac{\boldsymbol{g}(Z_{k+1}|\mathbf{X}_{k+1})\boldsymbol{\pi}_{k+1|k}(\mathbf{X}_{k+1}|Z_{1:k})}{\int \boldsymbol{g}(Z_{k+1}|\mathbf{X})\boldsymbol{\pi}_{k+1|k}(\mathbf{X}|Z_{1:k})\delta\mathbf{X}} \label{eq:FISST_update}
\end{align}
where $Z_{1:k}$ denotes the history of measurement data from time $1$ to $k$, $\boldsymbol{\pi}_{k+1|k}(\cdot|Z_{1:k})$  is  a multi-object predicted density, $\boldsymbol{\pi}_{k+1}(\cdot|Z_{1:k+1})$ is a multi-object filtering density, $\Phi_{k+1|k}(\cdot|\cdot)$  is a multi-object transition density, and $\boldsymbol{g}(\cdot|\cdot)$ is  a multi-object likelihood function. Notably, the integrals in Equations~\eqref{eq:FISST_predit}-\eqref{eq:FISST_update} are FISST integrals expressed as:
\begin{equation}
    \int f(\mathbf{X})\delta\mathbf{X} = \sum_{n=0}^{\infty}\frac{1}{n!}\sum_{(\ell_1,\dots,\ell_n) \in \mathbb{L}^n}\int_{\mathbb{X}^{n}} f(\{(x_{1},\ell_1), \ldots, (x_{n},\ell_{n})\})d(x_{1}, \ldots, x_{n}).
\end{equation}

\noindent However, in general, the exact computation of the FISST Bayes multi-object recursion is intractable, but the LMB density can be efficiently approximated via the LMB filtering recursion in~(Section III, \cite{reuter2014lmb}).

\subsection{Clustering Algorithms} \label{sec:clustering_algo}
Clustering involves partitioning a set of objects into groups or clusters, ensuring each group is comprised of objects with similar properties. Given the inherent similarity of this objective to that of the track-to-track fusion problem, it is intuitively sensible to harness the capabilities of advances in efficient clustering algorithms to address the track-to-track association challenge in DMOT.

Clustering algorithms can be broadly categorised into two types: i)~hierarchical, and ii)~partitioning clustering. Hierarchical algorithms build a nested hierarchy of clusters, organising data in the form of a tree. Classical hierarchical clustering algorithms are sensitive to noise and outliers and suffer from high computational complexity, limiting their application scenario in large-scale data sets~\cite{cluster_survey}. Several hierarchical techniques have been developed to address these limitations, for example, CURE~\cite{cure}, ROCK~\cite{rock} and BIRCH~\cite{birch}. However, in DMOT context, the hierarchical structure is generally unnecessary for achieving kinematic and label consensus.
In contrast, partitioning clustering algorithms divide the input data into distinct, non-overlapping subsets or clusters. Although numerous clustering algorithms, such as $k$-means~\cite{MacQueen1967}, EM-clustering~\cite{EM1977}, DBSCAN~\cite{ester1996density} , and Mean-Shift~\cite{fukunaga1975}, have been proposed to cater to various data structures, not all of them are suitable in a DMOT context. For instance, the $k$-means algorithm \cite{MacQueen1967} necessitates prior knowledge of the number of clusters, a requirement often unmet in sensor networks. 
In contrast, DBSCAN and Mean-Shift, are widely used and efficient clustering algorithms that do not necessitate prior knowledge of the number of clusters. However, their performance in problems is sensitive to the selection of a hyper-parameter---kernel bandwidth for Mean-Shift and the radius of the cluster neighbourhood $\epsilon$ for DBSCAN~\cite{rodriguez2014clustering}. Determining a suitable hyper-parameter depends on the tracking scenario and, in general, is a non-trivial task~\cite{cheng1995mean}. 
While early attempts are successful in determining $\epsilon$ in an online manner in a scenario with multiple sensors monitoring a common area (overlapping FoV)~\cite{dbscan_cutoff}, it cannot be directly extended to general DMOT problems where each sensor node may have distinct, partially overlapping FoVs.
In our work, we investigate a more generalisable alternative for DMOT problems. We begin with a problem formulation in the following section, while the subsequent sections elaborate on our approach.

\section{Distributed Track-to-Track Fusion with Clustering} \label{sec:proposed_method}
We introduce the new track-to-track fusion method to achieve kinematic and label consensus across a network of nodes with limited FoV sensors. In particular, we frame the problem under practical settings of sensor heterogeneity, unknown FoVs, limited computational capabilities of nodes, and communication bandwidth limits on communication channels.

We study fusing labelled estimates instead of densities to reduce the quantity of data shared over bandwidth-limited communication links. Importantly, the approach is agnostic to the local sensor modality and tracker. We intentionally avoid solving the challenging assignment problem, considering the limited processing capability of nodes. 
Instead, we consider modifications to adapt a recent advance in cluster analysis to group track estimates for fusion more effectively and efficiently to achieve kinematic consensus. The resulting approach is a generalisable cluster analysis method for DMOT problems because our proposal requires neither prior knowledge of the number of clusters nor the optimal selection of hyper-parameters.

Importantly, DMOT entails the estimation of multi-object trajectories, necessitating consistent labels or an SIAP among sensor nodes to address the label consistency problem. We introduce a new label algorithm to maintain a record of association history between track labels from different nodes to ultimately facilitate label consensus. In particular, the approach offers enhanced robustness against outlier labels.

We begin with a formal description of the problem (Section~\ref{sec:prob-description}) and develop the track association method along with the fusion of tracks for kinematic consensus (Section~\ref{sec:clustering-kinematic-consensus}), followed by our proposed label consensus algorithm in the following sections.

\begin{figure}[!h]
\centering \includegraphics[width=0.55\textwidth]{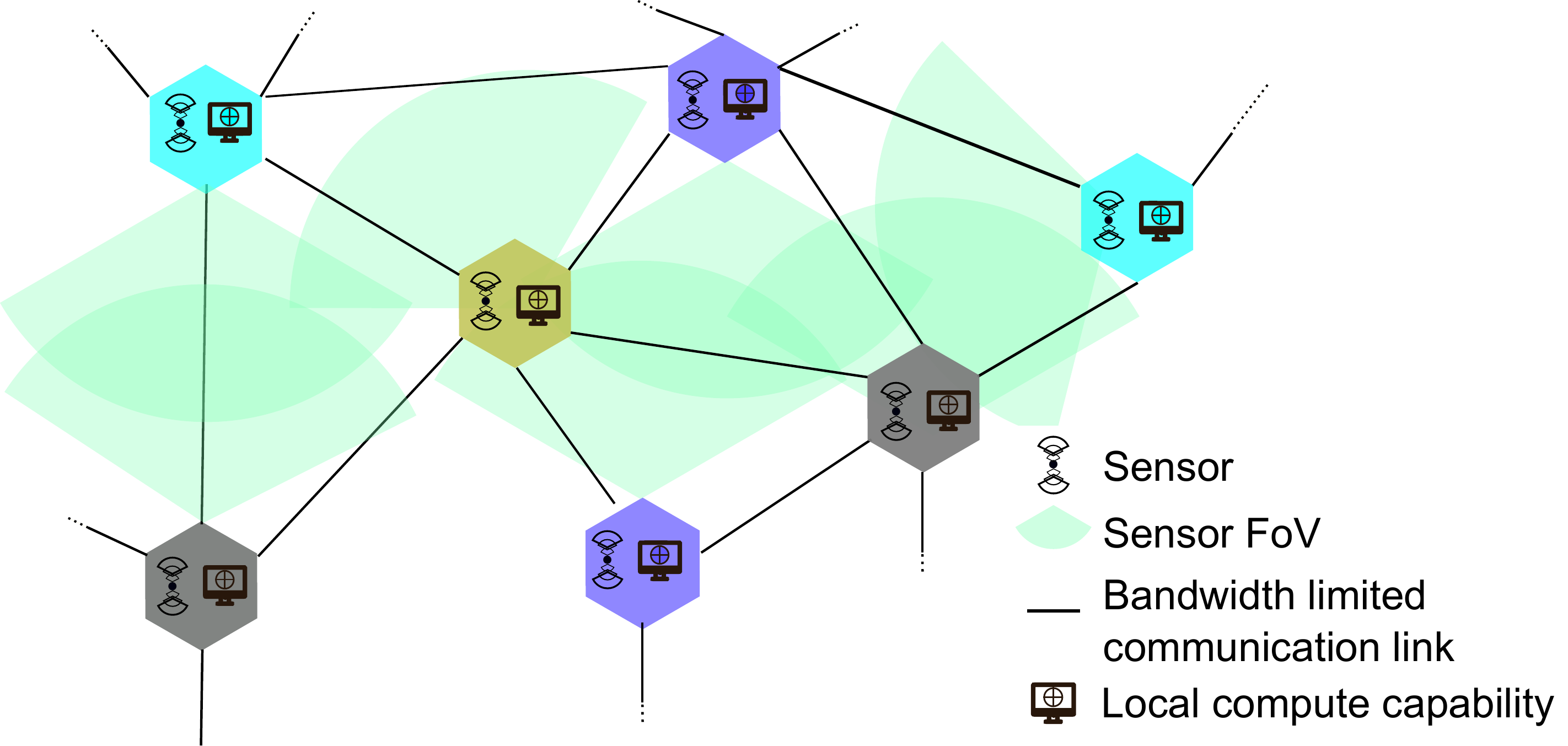}
\vspace{-0.3cm}
\caption{A distributed heterogeneous sensor network with limited FoVs and bandwidth-limited communication links.}
\label{fig:network_model} 
\end{figure}

\subsection{Problem Description}\label{sec:prob-description}
Consider a distributed sensor network, as illustrated in Figure~\ref{fig:network_model}, characterised by an undirected graph $(\mathcal{N},\mathcal{A})$ where $\mathcal{N}$ denotes the set of network nodes and $\mathcal{A} \in\mathcal{N}\times\mathcal{N}$ denotes the set of arcs corresponding to connections amongst nodes. The network performs the surveillance task of observing and tracking an unknown and time-varying number of mobile objects over time in a large region. Each node is equipped with a noisy, limited field-of-view sensor subject to false alarms and misdetections. Additionally, each node has a local computing unit for computing local multi-object state estimates. There are no central fusion nodes in the considered network. Further, each node is assumed to have a transceiver to send and receive multi-object state estimates, to and from, other nodes directly or via multi-hop routing, using a typical ad-hoc network or mesh network.

Instead of fusing local multi-object densities as in several RFS-based DMOT works~\cite{li2020Guchong,li2019computationally}, we propose fusing labelled multi-object state estimates to achieve consensus among all local nodes in $\mathcal{N}$. In particular, at time step $k$, every local node $n \in \mathcal{N}$ transmits a message $(n,\mathbf{X}_k)$ comprising both the node's identity $n$ and its local labelled multi-object state estimate $\mathbf{X}_k$. For brevity, we denote $\mathbb{I}=\mathbb{L}\times\mathcal{N}$ as the global label space of the sensor network, ensuring that each global label is unique throughout the entire network, and $N=|\mathcal{N}|$ denotes the number of nodes. Let $\mathbf{X}_k^{(n)}=\{(x,(\ell,n)):(x,\ell)\in\mathbf{X}_k\}\subset\mathbb{X}\times\mathbb{I}$ represent the globally labelled multi-object state estimate of node $n$ at time $k$, derived from the received message $(n,\mathbf{X}_k)$. Additionally, we define $\mathbf{L}_k^{(n)}=\mathcal{L}(\mathbf{X}_k^{(n)})=\{\ell^{(n)}\triangleq(\ell,n):\ell\in\mathcal{L}(\mathbf{X}_k)\}\subset\mathbb{I}$ as the global labels for estimate $\mathbf{X}_{k}^{(n)}$ and $X_{k}^{(n)} = \mathcal{S}(\mathbf{X}_k^{(n)})=\{\mathcal{S}(\mathbf{x}): \mathbf{x}\in\mathbf{X}_k^{(n)}\}$ as the set of non-labelled states where $\mathcal{S}:\mathbb{X}\times\mathbb{I}\rightarrow\mathbb{X}$ is the state extraction function given by $\mathcal{S}(x, \ell)=x$. Here $\ell^{(n)} = (s,\alpha,n) \in \mathbb{I}$ is the object label computed at node $n$, containing its birth time $s \leq k$, while $\alpha$ is a non-negative integer used to distinguish multiple objects born at the same time $s$.
We aim to fuse the local multi-object estimates $\{\mathbf{X}^{(n)}_k \}_{n \in \mathcal{N}}$ from $N$ connected nodes to achieve i) kinematic consensus, and ii) label consensus or a SIAP~\cite{dunham2004multiple} of the global multi-object state estimates $\mathbf{X}^{(\text{global})}_k$ across all nodes.

\subsection{Clustering Analysis to Find Density Peaks for Kinematic Consensus}
\label{sec:clustering-kinematic-consensus}

Solving the track-to-track fusion problem at any node involves fusing labelled multi-object state estimates. 
However, in general, fusing multi-object states among $N$ nodes is still burdened by a multi-dimensional assignment problem of $N$ dimensions. Recall that solving the optimal $N$-dimensional assignment problem for $N$ > 2 is known to be NP-hard and, thus, extremely computationally expensive. In this paper, we propose exploiting the efficiency of the clustering algorithm to tackle the difficulty, which leads to a more efficient (sub-optimal) solution to reduce computational and communication bandwidth demands.

We consider the recent clustering by fast search and find of density peak algorithm (CDP) \cite{rodriguez2014clustering} to analyse the state space of estimates to ultimately achieve kinematic consensus.
The CDP algorithm is a partitioned clustering method that clusters based on the density of data points.
Provided that two criteria are met: i) cluster centres are surrounded by neighbours with lower local density;
ii) the distance from a cluster centre to other points with higher local density is relatively large, the CDP algorithm has demonstrated good performance in challenging problems to identify cluster centres compared to other commonly used clustering algorithms~~\cite{wiwie2015comparing}. In DMOT settings, if each local sensor node employs a ``good'' multi-object tracker (e.g., MHT \cite{reid1979algorithm}, GLMB \cite{vo2013glmb}, or LMB \cite{reuter2014lmb}), it is reasonable to assume that estimates corresponding to the same object are closely located and therefore making CDP algorithm suitable for tracking scenarios.

Specifically, the CDP algorithm performs the following mapping: $\mathcal{M}: \mathbf{x}_i \rightarrow (\rho_i, \delta_i)$, where $\mathbf{x}_i \in \mathbf{X}^{(\text{local})}$ and $\mathbf{X}^{(\text{local})} = \underset{n\in\mathcal{N}}{\uplus}\mathbf{X}^{(n)}$ is the set of non-fused local state estimates available at a sensor node, $\uplus$ denotes the disjoint union of sets. This mapping transforms each input datum into a new $(\rho,\delta)$ coordinate, where the cluster centres are inherently distinguished from neighbouring centres.
Let $d(\mathbf{x}_{i}, \mathbf{x}_{j})$ be any arbitrary metric between $\mathbf{x}_{i}$ and $\mathbf{x}_{j}$. Then
\begin{align}
    \rho_{i} =& \sum_{j}\chi\left(d(\mathbf{x}_{i}, \mathbf{x}_{j}) - \xi_{c}\right)\label{eq:CDP_local_density}\\
    \delta_{i} =& \underset{j: \rho_{i} > \rho_{j}}{\min}~d(\mathbf{x}_{i}, \mathbf{x}_{j}) \label{eq:CDP_distance} 
\end{align}
where $\chi(c) = 1$ if $c<0$ and $\chi(c) = 0$ otherwise, and $\xi_{c}$ is the cutoff distance.
Here, $\rho_i$ can be regarded as a CDP local density, while $\delta_i$ can be regarded as a CDP density distance.

\begin{figure}[!tb]
\centering \includegraphics[width=0.6\textwidth]{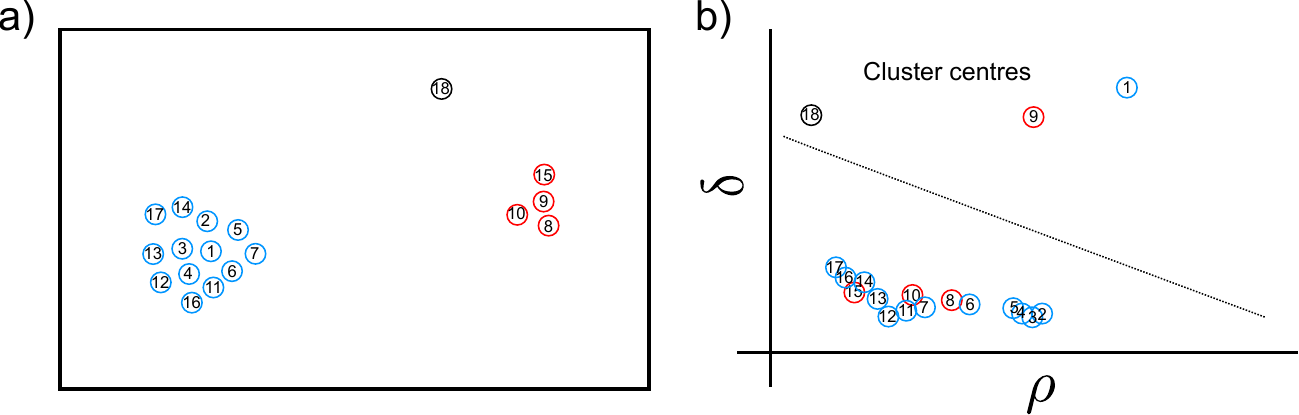}
\vspace{-0.3cm}
\caption{CDP algorithm illustration. Different colours denote different clusters of data. (a) Example set of data points. (b) Data points are converted to ($\rho,\delta$) coordinates. Cluster centres can be identified by data with high $(\rho, \delta)$ values (those above the dashed line in our example).}
\label{fig:CDP_sample} 
\end{figure}

The cutoff distance $\xi_c$ for the CDP algorithm is dynamically chosen such that the average number of neighbours for each cluster centre constitutes approximately $1\%$ to $2\%$ of the total number of points, as recommended in \cite{rodriguez2014clustering}. The CDP algorithm's cutoff distance $\xi_c$ is dynamically determined by estimating the density using a Gaussian kernel \cite{meanshift1995}. 
In DMOT settings, local estimates from the same sensor are guaranteed to originate from different objects. Hence, for any estimates that originate from the same local sensor, \ie{} $\mathbf{x}_{i}\in\mathbf{X^{(n)}}$ and $\mathbf{x}_{j}\in\mathbf{X^{(n)}}$, their distance $d(\mathbf{x}_{i}, \mathbf{x}_{j})$ is set to infinity to enforce this constraint.
Once all input data have been transformed, the cluster centres can be identified as the points for which its distance value $\delta$ is anomalously large, as illustrated in Figure~\ref{fig:CDP_sample}.

\begin{figure}[!htb]
\vspace{-0.3cm}
\centering \includegraphics[width=0.65\textwidth]{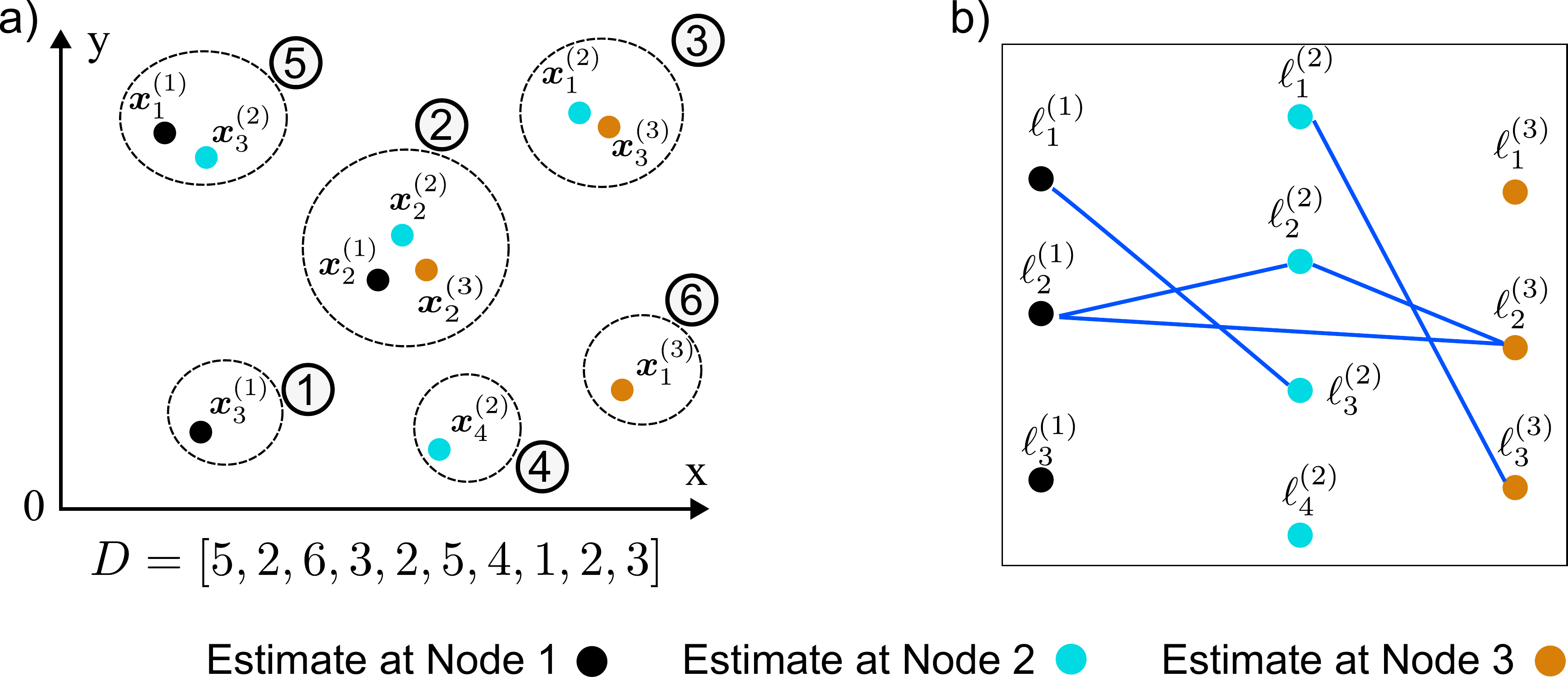}
\vspace{-0.3cm}
\caption{An example illustrating the analysis of labelled estimates with the \textsf{ModifiedCDP} Algorithm, resulting in label associations from the cluster analysis and the derivation of the kinematic consensus state. \textbf{a)} Consider ten 2D spatial locations of labelled local estimates from three sensor nodes: 
$\boldsymbol{X}^{(\text{local})} = [\boldsymbol{x}_1^{(1)},\boldsymbol{x}_2^{(1)},\boldsymbol{x}_3^{(1)},\boldsymbol{x}_1^{(2)},\boldsymbol{x}_2^{(2)},\boldsymbol{x}_3^{(2)},\boldsymbol{x}_4^{(2)},\boldsymbol{x}_1^{(3)},\boldsymbol{x}_2^{(3)},\boldsymbol{x}_3^{(3)}]$.
After applying the \textsf{ModifiedCDP} Algorithm, $6$ cluster centres are identified. Now, the resulting clustering index vector for the input $\boldsymbol{X}^{(\text{local})}$ is $D = [5,2,1,3,2,5,4,6,2,3]$. This vector signifies that $\boldsymbol{x}^{(1)}_1$ is associated with $\boldsymbol{x}^{(2)}_3$ (clustering index $5$), $\boldsymbol{x}^{(1)}_2$ is associated with both $\boldsymbol{x}^{(2)}_2$ and $\boldsymbol{x}^{(3)}_2$ (clustering index $2$), $\boldsymbol{x}^{(2)}_1$ is associated with $\boldsymbol{x}^{(3)}_3$ (clustering index $3$), whilst the remaining local estimates are not associated with any others.
\textbf{b)} Illustrates the label associations between local estimates based on the clustering indices $D$ obtained from the \textsf{ModifiedCDP} Algorithm. This information is used to update the weighted graph to achieve global label consensus. 
}
\label{fig:clustering_example} 
\end{figure}

The cluster centres are identified by selecting ($\rho,\delta$) points that are above density threshold $\tau_{\rho}$ and distance threshold $\tau_{\delta}$ through a manual inspection.
Intuitively, the CDP algorithm naturally separates cluster centres from their neighbours into two clusters in the $(\rho,\delta)$ coordinate space. Therefore, we propose automating the cluster distillation process using the $k$-means clustering algorithm~\cite{MacQueen1967} with $k=2$ to separate cluster centres and their neighbours. This allows us to adopt the CDP algorithm to analyse the labelled state estimates. 
Now, cluster centres are identified as the points with relatively high $(\rho, \delta)$ values. Subsequently, all remaining points are assigned to their nearest cluster centre. Algorithm~\ref{algo:modified_CDP} summarises the modified CDP algorithm for analysing the set of local and shared labelled estimates. Figure~\ref{fig:clustering_example}(a) provides an example outcome of an analysis of labelled estimates at a node with the \textsf{ModifiedCDP} Algorithm to tackle the complexity of the multi-dimensional assignment problem for achieving kinematic consensus.

\begin{figure}[tb!]
    \centering
    \begin{minipage}{.7\linewidth}

    \begin{algorithm}[H]
        {\footnotesize{}{}{}\caption{\textsf{ModifiedCDP}}
        \label{algo:modified_CDP}
        \begin{algorithmic}[1]  		
        \State \textbf{Input}: 
        $\boldsymbol{X}^{(\text{local})} = \big[\boldsymbol{x}_1^{(1)},\dots,\boldsymbol{x}_{M_1}^{(1)},\dots, \boldsymbol{x}^{(N)}_1,\dots,\boldsymbol{x}^{(N)}_{M_N}\big]$ 
        \State \textbf{Output}: Cluster Index vector $D$
        \State $D := \operatorname{zeros}(|\boldsymbol{X}^{(\text{local})}|)$
        \State $\operatorname{ID} := 0$

        \LComment{Cluster analysis of the state space}
        \State $\rho := $ Compute CDP density of $\boldsymbol{X}^{(\text{local})}$ using \eqref{eq:CDP_local_density}
        \State $\delta := $ Compute CDP distance of $\boldsymbol{X}^{(\text{local})}$ using \eqref{eq:CDP_distance} 
        \State $ClusterCentreSet:=$ Select high density estimates $(\rho, \delta)$ with $k$-means ($k$=2)

        \LComment{Assign each cluster centre a unique ID} 
        \For{each $\boldsymbol{x}_{i}\in\boldsymbol{X}^{(\text{local})}$}
            \If{$\boldsymbol{x}_{i}\in ClusterCentreSet$}      
                \State $\operatorname{ID} := \operatorname{ID} + 1$
                \State $D[i] := \operatorname{ID}$
            \EndIf
        \EndFor
        
        \LComment{Assign estimates to cluster centres using metric $d$}
        \For{each $\boldsymbol{x}_{i}\in\boldsymbol{X}^{(\text{local})}$} 
            \If{$\boldsymbol{x}_{i}\notin ClusterCentreSet$}      
                \State $j := \underset{k}{\operatorname{argmin}}~d(\boldsymbol{x}_{i}, \boldsymbol{x}_{k}) ~\text{where}~  \boldsymbol{x}_{k}\in ClusterCentreSet$ 
                \State $D[i] := j$
            \EndIf
        \EndFor
        \end{algorithmic}}
    \end{algorithm}
\end{minipage}
\end{figure}

\begin{remark}
The pairwise metric $d(\mathbf{x}_{i}, \mathbf{x}_{j})$ used in \eqref{eq:CDP_local_density} and \eqref{eq:CDP_distance} between $\mathbf{x}_i$ and $\mathbf{x}_j$ may be represented as a Euclidean distance, utilising solely the contemporaneous local estimates $\mathbf{x}_i$ and $\mathbf{x}_j$  corresponding to a track length of $1$. This can be readily extended to incorporate local track estimates of arbitrary lengths through the employment of the OSPA\textsuperscript{(2)} metric~\cite{beard2020a} or alternative track-to-track metrics. 
\end{remark}

\subsection{Weighted Label Graph for Global Label Consensus}
\label{sec:wgl} 
MOT concerns both object positions (kinematics) and identities (labels). Although the problem of kinematic consensus is addressed under limited FoVs of multiple nodes, the problem of achieving label consensus to realise an SIAP across the distributed network~\cite{hoa2021distributed} remains. In other words, if node $1$ assigns the label $\ell_i$ to the true object $i$, then our desire is for all other nodes in the network to assign the label $\ell_i$ to the true object $i$. For a formal definition of label consensus, we direct the reader to Definition~3 in ~\cite{hoa2021distributed}.

Recall that during kinematic consensus, all labelled state estimates are included in the fused multi-object state estimates; this can include labelled estimates from clutter. Further, clustering analysis for kinematic consensus may incorrectly associate state estimates from different objects. Therefore, a carefully designed algorithm is needed to resolve inconsistencies and achieve label consensus. In this section, we present our algorithm formulation for achieving label consensus across a network of limited FoV sensor nodes.

Since labels are discrete, an undirected graph is a natural representation of the association among labels. However, naive use of a graph-based algorithm can result in incorrect label consensus due to labels resulting from false-alarm estimates and incorrect associations from clustering analysis of labelled estimates at a given node. In this work, we propose using a weighted graph to represent the associations among local labels derived from the cluster indices $D$.  Specifically, let $G=(V,E, w)$ be a graph, where:
\begin{itemize}
    \item $V$ represents the set of vertices of $G$, comprising all the local labels, \ie, $V = \uplus_{i=1}^k \uplus_{n=1}^N \boldsymbol{L}^{(n)}_i $, where $\boldsymbol{L}^{(n)}_i$ denotes the local labels of a node $n$ at time $i$. 
    \item $ E\subseteq \big\{ \{\ell_i,\ell_j\}\mid \ell_i,\ell_j \in V\;{\textrm {and}}\; \ell_i\neq \ell_j\big\}$ constitutes the set of edges of $G$. Each edge $e = \{ \ell_i, \ell_j \} \in E$ signifies  that $\ell_i$ is associated with $\ell_j$. 
    \item $w$ denotes the edge weight assigned to each edge $e \in E$. Here, the edge weight represents a \textit{distance} between two vertices of label graph $G$ to capture the likelihood of two distinct labels corresponding to the same underlying object; a smaller value indicates a higher likelihood. 
\end{itemize}

\noindent Given the graph representation, our goals are to: 
\begin{enumerate}
    \item Build and maintain the label associations across the network of nodes from the information of labelled estimates shared across the network.
    \item Employ the historical information in graph $G$ to assign global fusion labels to achieve label consensus across the sensor network.
\end{enumerate}
This inevitably requires formulating a graph update method to incorporate new information into an existing label graph $G$. Our problem formulation is based on minimising shared information, thus we are limited to updating a graph $G$ at each local node from information derived from state estimates. We develop a method to incorporate new label associations resulting from clustering analysis into the association history. The graph updating algorithm we developed, \textsf{UpdateGraph} detailed in Algorithm~\ref{algo:update_graph},  uses the cluster indices $D$ generated by Algorithm~\ref{algo:modified_CDP} 
and local labels shared by nodes to update the historical information of label associations.

\begin{figure}[tb!]
    \centering
    \begin{minipage}{.7\linewidth}
    \begin{algorithm}[H]
        {\footnotesize{}{}{}\caption{\textsf{UpdateGraph}}
        \label{algo:update_graph}
        \begin{algorithmic}[1]  		
        \State \textbf{Input}: $G=(V,E,w), D, \boldsymbol{L}^{(\text{local})}$ \Comment{Current graph with Vertices, Edges and weights, clustering indices, local labels of all nodes}
        
        \State \textbf{Output}: $G$ \Comment{the updated graph} 
        \For{each $\ell \in \boldsymbol{L}^{(\text{local})}$} 
            \If{$\ell \not\in V$}
            \State $V := V \uplus \ell$ \Comment{add a new label as a vertex}
            \EndIf
        \EndFor
        \For{$m=1 \text{ to } \max(D)$}
            \State $\boldsymbol{L}_m^{(\text{local})} := $ local labels sets with cluster index $m$
            \LComment{Check all unique pairwise combinations of labels}
            \For{$\{\ell_{i}, \ell_{j}\} \in \mathcal{F}_2(\boldsymbol{L}_m^{(\text{local})})$}
                \If{$\{\ell_i,\ell_j\} \not\in E$}
                    \State $E := E \uplus \{\ell_i,\ell_j\} $ \Comment{add $\{\ell_i,\ell_j\}$ as a new edge }
                    \State $w(\{\ell_i,\ell_j\}) := w_{\max}$ \Comment{initialise a new edge weight}
                \Else
                    \State $w(\{\ell_i,\ell_j\}) := \max(0,w(\{\ell_i,\ell_j\})-1)$ \Comment{update an existing edge weight}
                \EndIf
            \EndFor
        \EndFor
        \State $G:=(V,E,w)$ \Comment{update $G$ from updated $V,E,w$ values}
        \end{algorithmic}}
    \end{algorithm}
\end{minipage}
\end{figure}

Specifically, in \textsf{UpdateGraph}, new labels are added as new vertices of the graph $G$ (lines $3-5$). For each cluster index $m \in D$, any two distinct labels, i.e., $\ell_i,\ell_j \in \boldsymbol{L}^{(\text{local})}_m$ with $\ell_i \neq \ell_j$, an edge $e = \{\ell_i,\ell_j\}$ with weight $w(e) = w_{\max}$ is added to $E$ if $e \not\in E$ (see lines $10-11$). Conversely, if $e \in E$, the weight of this edge $e$ is updated as $w(e) = \max(0, w(e) -1)$ (refer to line $13$). The maximising operator ensures that the weight of each edge remains non-negative.

Figure~\ref{fig:clustering_example} provides an example outcome of an analysis of labelled estimates at a node with the \textsf{ModifiedCDP} Algorithm and the resulting label associations from the analysis. Our \textsf{UpdateGraph} Algorithm utilises the derived associations from this process to update the graph $G$ and capture label associations over time. 

Once the label graph $G$ is updated, we can reassign labels to current fused estimates (kinematic consensus state) to realise global label consensus. The representation and update of the weighted label graph facilitate the formulation of a robust method for global label consensus as explained in Section~\ref{sec:dmot_fusion} and Algorithm~\ref{algo:compute_glob_est}.

\begin{figure}[htb!]
    \centering
    \begin{minipage}{.75\linewidth}
    \begin{algorithm}[H]
        {\footnotesize{}{}{}\caption{\textsf{DMOTFusion}}
        \label{algo:compute_glob_est}
        \begin{algorithmic}[1]  		
        \State \textbf{Input}: $\boldsymbol{X}^{(\text{local})},G$ \Comment{local estimates of all nodes, weighted label graph} 
        \State \textbf{Output}: $\boldsymbol{X}^{(\text{global})},G$ \Comment{global consensed labelled multi-object estimates  } 
    
        \State $D:=\textsf{ModifiedCDP}(\boldsymbol{X}^{(\text{local})})$ \Comment{Clustering to find common estimates}
        \State $\boldsymbol{L}^{(\text{local})}:=\mathcal{L}(\boldsymbol{X}^{(\text{local})}); X^{(\text{local})}:=\mathcal{S}(\boldsymbol{X}^{(\text{local})}); $ \Comment{extract local labels and local kinematic estimates}
        \State $G:=\textsf{UpdateGraph}(G,D,\boldsymbol{L}^{(\text{local})})$; \Comment{update graph with the new clustering indices}

        \State $\boldsymbol{X}^{(\text{global})} := \emptyset$ \Comment{initialise the global estimates}
        \For{$m=1 \text{ to } \max(D)$}
            \Statex \textit{\quad//Kinematic consensus}
            \State Compute $x^{(\text{global})}_m$ via $X_m^{(\text{local})}$ using  \eqref{eq:x_fused}
            \Statex \textit{//Global label consensus}
            \State $\ell_{m,\min} := \min(\boldsymbol{L}_m^{(\text{local})})$ \Comment{select the smallest label via lexicographical order}
            \State $\boldsymbol{L}_{G}:=\textrm{nearest}(G, \ell_{m,\min})$ \Comment{extract all the nearest labels of $\ell_{m,\min}$ from graph $G$}
            \State $\ell^{(\text{global})}_m := \min(\boldsymbol{L}_{G})$ \Comment{select the smallest label via lexicographical order}
            \Statex \textit{//Consensed globally labelled state estimates or tracks}
            \State $\boldsymbol{X}^{(\text{global})} := \boldsymbol{X}^{(\text{global})} \uplus \big( x^{(\text{global})}_m, \ell^{(\text{global})}_m\big)$
        \EndFor

        \end{algorithmic}}
    \end{algorithm}
\end{minipage}
\end{figure}

\begin{figure}[!tbh]
\vspace{-0.3cm}
\centering \includegraphics[width=0.6\textwidth]{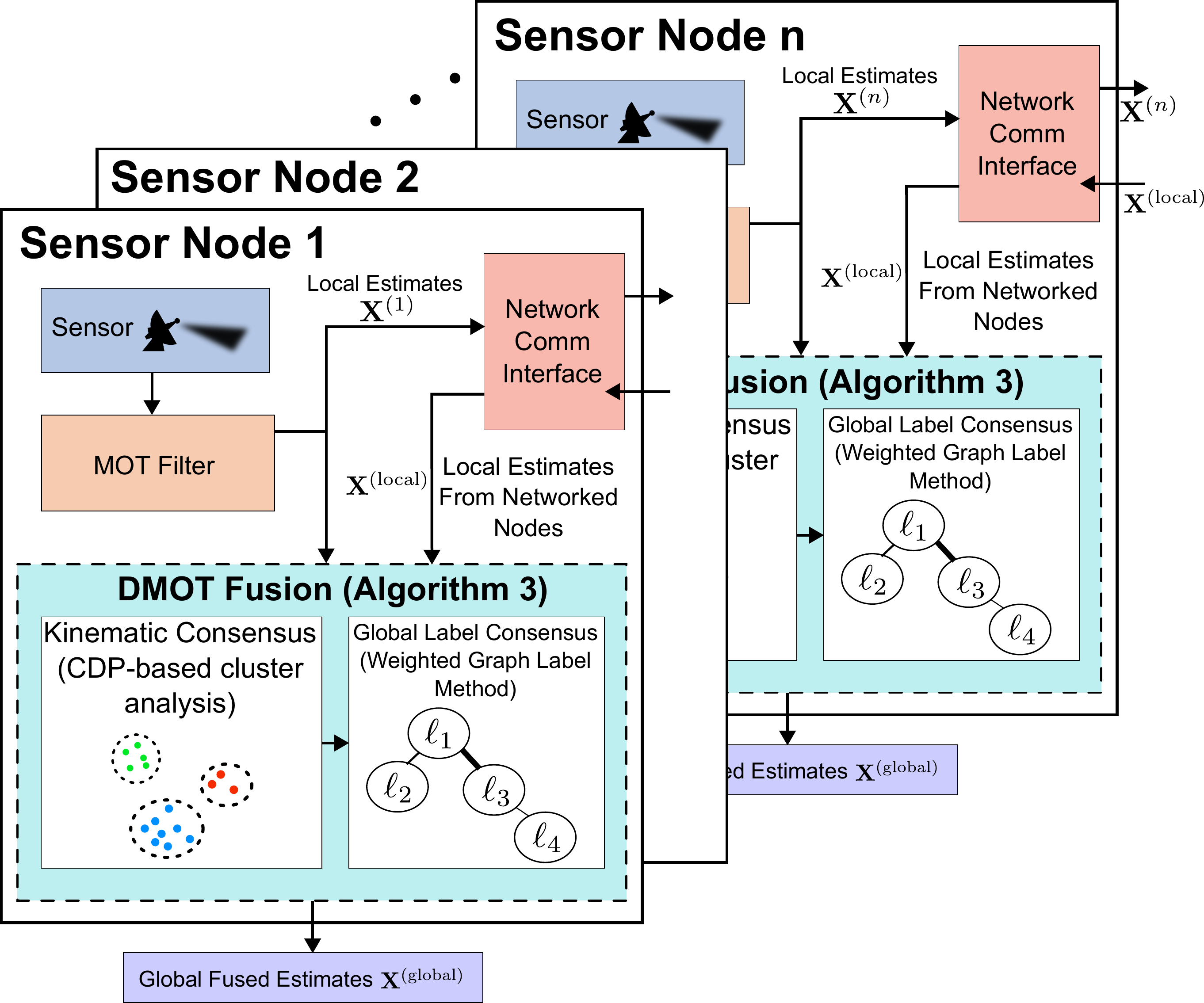}
\vspace{-0.3cm}
\caption{An overview of the proposed \cdp{} algorithm for Distributed MOT (DMOT). Each Sensor Node tracks objects within its sensor's FoV using a local MOT Filter and shares the local labelled estimates $\mathbf{X}^{(n)}$ via a communication network with other nodes. Global fused, labelled estimates $\mathbf{X}^{(\text{global})}$ or \textit{tracks} are computed at each sensor node using the local and shared information using \textsf{DMOTFusion} described in Algorithm~\ref{algo:compute_glob_est} where: i)~Kinematic Consensus is supported by the proposed CDP cluster analysis based algorithm, \textsf{ModifiedCDP}, described in Algorithm~\ref{algo:modified_CDP}; and ii)~Global Label Consensus is supported by the weighted graph labelling method using \textsf{UpdateGraph} described in Algorithm~\ref{algo:update_graph}.}
\label{fig:framework} 
\end{figure}

\subsection{DMOT Fusion}\label{sec:dmot_fusion}
Our complete distributed fusion algorithm \textsf{DMOTFusion} is summarised in Algorithm~\ref{algo:compute_glob_est}, and we provide an overview illustration of the proposed \cdp{} algorithm for DMOT in Figure~\ref{fig:framework}. Local sensor labelled estimates $\mathbf{X}^{(\text{local})}$ are first clustered via the CDP clustering algorithm (line $3$), and the label graph $G$ is updated according to the clustering result (line $5$).
Kinematic consensus is reached by fusing all local estimates that belong to the same cluster (in line $8$), and label consensus is achieved by assigning labels to fused estimates via the updated graph $G$ (in lines $9-11$). We describe: i)~the kinematic; and ii)~global label consensus methods, in detail, next.

In the absence of prior knowledge about local sensor capabilities, for kinematic consensus, the arithmetic average is given by: 
\begin{equation} \label{eq:x_fused}
    {x}^{(\text{global})}_m = \bigg[\sum_{{x} \in {X}_m^{(\text{local})}} {x}  \bigg]/|{X}_m^{(\text{local})}|,
\end{equation}
where ${X}_m^{(\text{local})}$ is a set of local state estimates from ${X}^{(\text{local})}$ with the same clustering index $m$, can be employed to fuse the estimates. For example, Figure~\ref{fig:clustering_example}(a) illustrates a  scenario in which our proposed cluster analysis with the \textsf{ModifiedCDP} Algorithm generates a proposal for associations between kinematic states. Here, the kinematic consensus state (fused global estimates) for the illustrated example can be computed by \eqref{eq:x_fused} to yield ${X}^{(\text{global})} = [{x}^{(1)}_3,{x}^{(\text{global})}_2,{x}^{(\text{global})}_3,{x}^{(2)}_4,{x}^{(\text{global})}_5,{x}^{(3)}_1]$.

\begin{remark}
Our approach is independent of the multi-object tracking techniques employed at a local node. In particular, we do not assume that the nodes share covariances along with the estimates. This is purposefully chosen to minimise bandwidth requirements on communication channels to suit practical settings. However, suppose the local tracking algorithm provides covariances. In that case, it is possible to fuse the means and covariances by, for example: i)~employing Mahalanobis distance between two Gaussian distributions for the distance metric $d$ in \eqref{eq:CDP_distance};  and ii)~computing the fused estimate in \eqref{eq:x_fused}  and its covariance employing an approach such as the GCI for multiple sensors or the optimal fusion method for two-sensors~\cite{barshalom1986the}.
\end{remark}

To derive the unique identity of consensus kinematic estimates and achieve global label consensus, given an updated graph $G$ generated from \textsf{DMOTFusion}, for each clustered index $m \in D$, the smallest label $\ell_{m,\min} = \min( \boldsymbol{L}^{(\text{local})}_m)$ in graph $G$ is chosen through a lexicographical order based on its birth time and node identity,~\ie:
\begin{align}
    \ell_i =  (s_i,\alpha_i, n_i) < \ell_j =  (s_j,\alpha_j, n_j) \Leftrightarrow s_i < s_j \text{ or } (s_i = s_j \text{ and } n_i < n_j). 
\end{align}

Subsequently, from the smallest local label $\ell_{m,\min}$, we find the set of label vertices that has a distance of $0$, \ie~$\mathbf{L}_{G} = \{\ell'_m\in V| d_{G}(\ell_{m,\min}, \ell'_m) = 0 \}$, where $d_{G}(v, u)$ is the distance between vertex $v$ and $u$ in graph $G$. The global label $\ell_m^{(\text{global})}$ is then selected as the smallest value from the set $\mathbf{L}_{G}$.

The basis for the consensus formulation is built on the key concept that: i)~the label graph represents the history of all previous label associations; and then ii) by assigning the smallest label (equivalent to the earliest object appearance) as the global consensus label to an object, prevents a larger label (later appearance) from being assigned to the same object and thus reduces label inconsistency.

\begin{figure}[!tb]
\centering \includegraphics[width=1\textwidth]{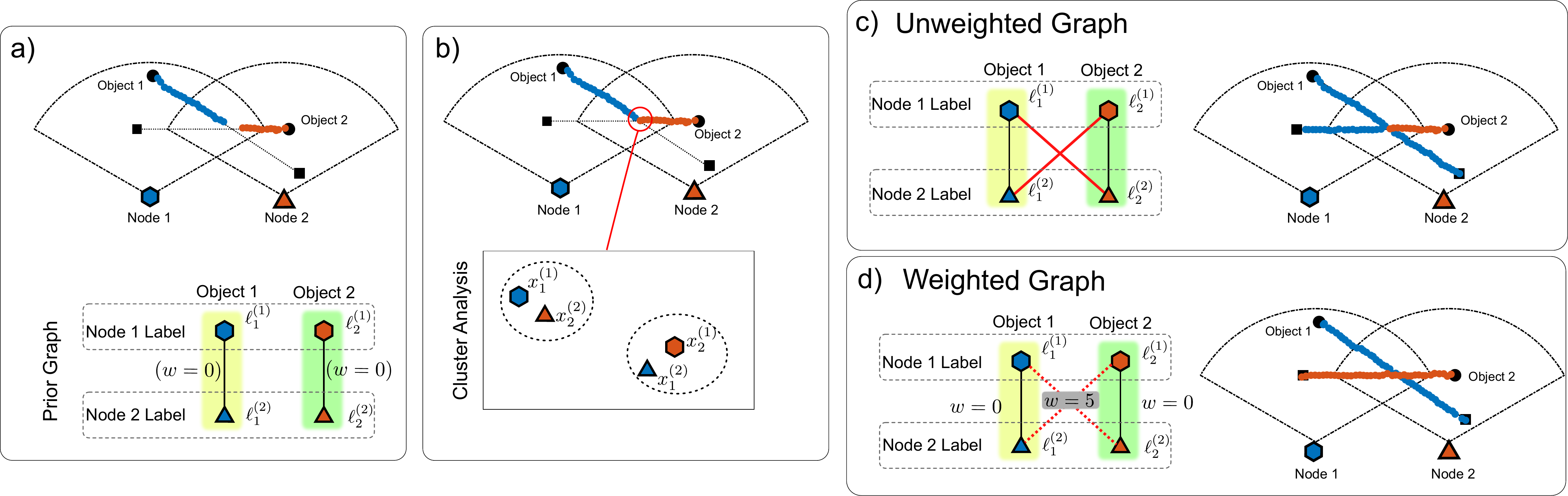}
\vspace{-0.7cm}
\caption{A simple illustration of the weighted graph label consensus method compared to the unweighted method in the presence of an incorrect label association assertion resulting from the clustering analysis. \textbf{a)} Object 1 and Object 2 enter the common FoV of sensor node 1 and node 2. The label graph shows the correct label association between each sensor node's labels. \textbf{b)} A clustering error occurs when object 1 crosses Object 2 at a similar space-time dimension and leads to grouping local estimates of Object~1 and 2 from nodes~1 and 2 into a single cluster. \textbf{c)} Result of unweighted graph label consensus method. After the label graph update, new edges $\{\ell_{1}^{(1)}, \ell_{2}^{(2)}\}$ and $\{\ell_{2}^{(1)}, \ell_{1}^{(2)}\}$ are added as a result of the clustering error. The smallest label $\ell_{1}^{(1)}$ in terms of lexicographic order is then permanently associated with Object~2 and causes a label switch while object 1's label is not affected. \textbf{d)} Effect of weighted graph label consensus method. After the graph update, the same edges are added with an initial weight of $5$ while all other edges' weights remain at $0$ (due to the graph update prior to the cluster error). The edge weight provides evidence to support a label association, effectively isolating Object 1 and 2's label graphs and preventing a label switch for Object 2.}
\label{fig:wgl_example} 
\end{figure}

Importantly, compared with the unweighted graph label consensus algorithm in~\cite{hoa2021distributed}, the formulation of weights serves to confirm a label association and reject spurious track labels resulting from false tracks or incorrect clustering results. Consequently, the proposed formulation is expected to improve track label consistency. For example, without the introduction of weights, it is possible for a clustering analysis error to cause label switching since the global consensus label is chosen to be the smallest label in lexicographic order. Figure~\ref{fig:wgl_example} illustrates the problem in a simple scenario in which our proposed weighted graph-based approach achieves label consensus across the sensor network in contrast to an unweighted graph formulation. Effectively, the edge weight serves to provide the evidence to support the label association and reject a spurious clustering result. 

\begin{table}[!tb]
    \centering
    \small
    \caption{Time complexity of algorithms for solving the multi-dimensional assignment problem.}
    \label{tab:algo_complexity}
    \begin{tabular}{l|llll}
    \hline
    Algorithm      & TC-OSPA\textsuperscript{(2)}\cite{hoa2021distributed} & DBSCAN 
    &Mean-Shift & CDP \\ 
    \hline
    \hline
    Average time complexity & $\mathcal{O}(|\mathcal{N}||\mathbf{T}_{\max}|^{4})$ & $\mathcal{O}(|\mathcal{N}||\mathbf{T}_{\max}|\log(|\mathbf{T}_{\max}|))$ & $\mathcal{O}(|\mathcal{N}||\mathbf{T}_{\max}|^{2})$ & $\mathcal{O}(|\mathcal{N}||\mathbf{T}_{\max}|^{2})$\\
    Worst case time complexity & $\mathcal{O}(|\mathcal{N}||\mathbf{T}_{\max}|^{4})$ & $\mathcal{O}(|\mathcal{N}||\mathbf{T}_{\max}|^{2})$ & $\mathcal{O}(|\mathcal{N}||\mathbf{T}_{\max}|^{2})$ & $\mathcal{O}(|\mathcal{N}||\mathbf{T}_{\max}|^{2})$\\
    \hline
    \end{tabular}
\end{table}

\subsection{Time Complexity and Bandwidth Comparison}
The key advantages of our proposed approach for distributed MOT are low computational complexity and low bandwidth requirements. In this section, we analyse and compare the computational complexity and bandwidth demands of our method alongside other alternatives for DMOT.

Table~\ref{tab:algo_complexity} summarises the average and worst-case time complexities of the clustering-based analysis method and TC-OSPA\textsuperscript{(2)} \cite{hoa2021distributed} algorithms. Compared to the former DMOT algorithm, employing pair-wise assignment, the cluster analysis algorithms (our proposed method, CDP-based, and DBSCAN-based, used for comparison) are significantly less computationally demanding than the TC-OSPA\textsuperscript{(2)} algorithm.

We consider the bandwidth requirements for labelled estimate fusion, labelled estimate \& covariance fusion, and full LMB density fusion methods in a sensor network with $|\mathcal{N}|$ sensor nodes and $|\mathbf{T}_{\text{max}}|$ objects. The results of the analysis are summarised in Table~\ref{tab:bandwidth_table}. 

\begin{table}[!tb]
\centering
\small
\caption{Bandwidth requirement comparison of different information sharing strategies for fusion. Here, $n_{l}$ and $n_{d}$ are the track label dimension and kinematic state dimension, respectively, $\mu$ is the ratio between the number of Bernoulli components and confirmed objects in an LMB density.}
\label{tab:bandwidth_table}
\begin{tabular}{l|l}
\hline
Method      & Bandwidth (bytes) \\ 
\hline
\hline
Labelled Estimates &  $8 \cdot (n_{l} + n_{d})|\mathcal{N}||\mathbf{T}_{\max}|$ \\
Labelled Estimates \& Covariance & $8 \cdot (n_{l} + n_{d} + \frac{n_{d}(n_{d} + 1)}{2})|\mathcal{N}||\mathbf{T}_{\max}|$ \\
LMB Density & $8 \cdot  [1 + n_{l} + n_{d} + \frac{n_{d}(n_{d} + 1)}{2}]|\mathcal{N}||\mathbf{T}_{\max}|\mu$\\
\hline
\end{tabular}
\end{table}

The dimensions of each state estimate and label are $n_{d}$ and $n_{l}$, respectively. Each dimension is presumed to be represented by an 8-byte floating-point value. In density fusion, each LMB density can be uniquely defined by its labelled Bernoulli components $\{r(\ell),p(\cdot,\ell)\}_{\ell \in \mathbb{L}}$. The number of labelled Bernoulli components typically exceeds the number of objects $|\mathbf{T}_{\max}|$ and therefore the exact number of objects in a full LMB density is represented by $|\mathbf{T}_{\max}|\cdot\mu$ where $\mu \geq 1$.

Labelled estimates fusion is the most bandwidth-efficient method, only requiring each local sensor to share its estimates within the network. To enhance tracking accuracy, the covariance of each track can also be shared across the network, leading to additional bandwidth requirements. As expected, sharing the full multi-object density demands significantly high bandwidth; depending on the number of objects and sensors, the bandwidth demands can be prohibitively high for practical applications~\cite{hoa2021distributed}.

\section{Numerical Experiments} \label{sec:experiment}
In this section, we carry out a series of experiments using: i)~simulated data (Section~\ref{sec:hetero-exp}); and ii)~real-world trajectory data (Section~\ref{sec:taxi-trad-exp}) to assess the performance of our proposed DMOT strategies under networks of heterogeneous sensors where all local nodes operate using labelled filters.

We compare the proposed clustering-based multi-object tracking algorithm under limited FoV sensors with weighted graph labelling method using both our proposed cluster analysis with CDP together with DBSCAN and Mean-Shift as alternative cluster analysis baselines as \cdp{}, \dbscan{} and \meanshift{}, respectively. Then we also ablate our WGL method to illustrate its effectiveness and provide baselines with the three cluster analysis methods without the WGL method as \uncdp{}, \undbscan{} and \meanshift{}.

Notably, TC-OSPA\textsuperscript{(2)} represents the current state-of-the-art DMOT algorithm capable of addressing limited FoV sensors in \textit{labelled} trackers by performing a sequential pairwise matching between local estimates for track-to-track fusion. Therefore, we also compare with TC-OSPA\textsuperscript{(2)} in~\cite{hoa2021distributed}. Additionally, we implement the \textit{centralised} multi-sensor LMB (MS-LMB)~\cite{vo2019multisensor} to show the estimation error in an ideal case where all sensors' FoVs are known. We describe the experimental settings and performance evaluation measures we employed below.

\vspace{0.2cm}
\noindent\textbf{Experimental Settings.~}We employ an LMB filter with a Gaussian mixture implementation at each local sensor node. Gibbs sampling~\cite{vo2016efficient}, an efficient variant of the Metropolis-Hastings algorithm, is utilised for the joint prediction and update step to sample multiple association hypotheses efficiently. This approach balances computational efficiency and accuracy in handling complex MOT scenarios.  Alternatively, Murty's algorithm~\cite{vo2014glmb} can be employed for the deterministic computation of multiple association hypotheses. However, this method is significantly slower than Gibbs sampling, particularly in scenarios with a large number of objects or measurements.
The maximum number of hypotheses retained for each LMB filter is $1000$, and the existence pruning threshold is fixed at $10^{-5}$. An object is confirmed when its existence exceeds $0.5$. As the initial locations of the objects are unknown, we apply the adaptive birth procedure for the LMB filter~\cite{reuter2014lmb}.
Although our proposed CDP analysis for tracking problems has the benefit of avoiding parameter tuning, DBSCAN requires determining appropriate settings. The minimum number of points required to form a dense region, $minPts$ is set to 1 for the DBSCAN algorithm with the neighbourhood radius $\epsilon = \SI{30}{\meter}$ and $\epsilon=\SI{250}{\meter}$ set for the simulated trajectory experiments (Section~\ref{sec:hetero-exp}) and real-world trajectory experiments (Section~\ref{sec:taxi-trad-exp}), respectively, for optimal clustering performance.
Gaussian kernel bandwidth of $\SI{50}{\meter}$ is used for the Mean-Shift algorithm for all scenarios.

\vspace{0.2cm}
\noindent\textbf{Performance Evaluation.~}We evaluate the tracking performance using the following metrics: 
i) Optimal Sub-Pattern Assignment (OSPA)~\cite{schuhmacher2008consistent}, 
ii) OSPA-on-OSPA (OSPA\textsuperscript{(2)})~\cite{beard2020a} with cut-off $c=100$~m, parameter $p=1$, and window length $10$, and 
iii) the averaged fusing time of each measurement time step for the considered fusion algorithms. 
In particular, OSPA considers not only localisation error but also cardinality (i.e., number of objects) error, while OSPA\textsuperscript{(2)} takes into account localisation, cardinality, track fragmentation, and track label switching errors. A smaller metric value indicates better tracking performance. We refer to \cite{beard2020a} and the references therein for detailed computations of these metrics.  The reported results are averaged over $100$ Monte Carlo (MC) runs.

\subsection{Heterogeneous Sensor Network}\label{sec:hetero-exp}
\begin{figure}[!tb]
\centering \includegraphics[width=0.75\textwidth]{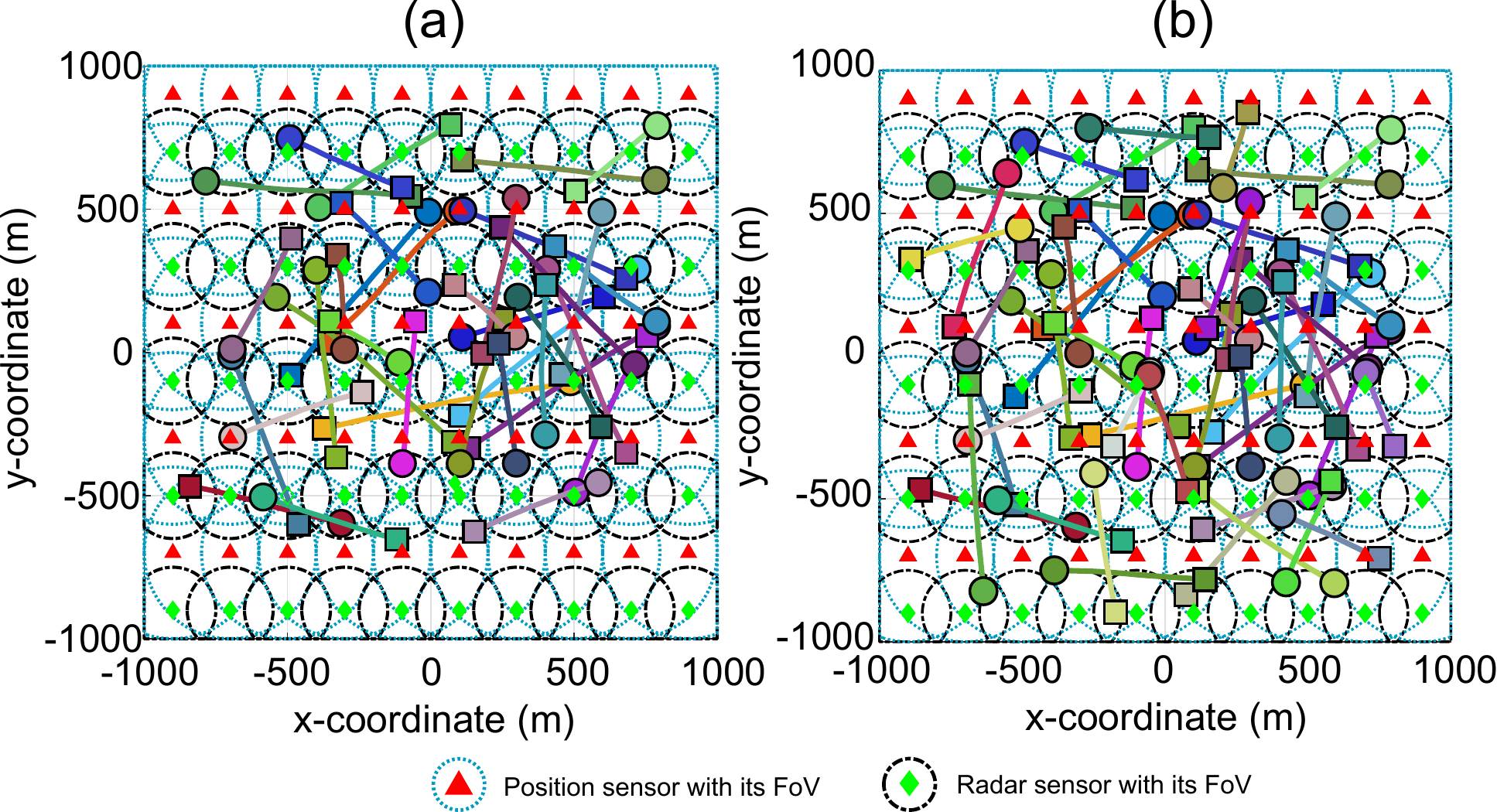}
\vspace{-0.3cm}
\caption{Heterogeneous sensors scenario: $100$ sensor nodes tracking (a) $35$ mobile objects; (b) $50$ mobile objects. Starting and stopping positions are denoted by $\ocircle$ and $\Box$.}
\vspace{-0.3cm}
\label{fig:s2_truth} 
\end{figure}

In this scenario, we consider the problem of using an interconnected network of $100$ heterogeneous sensors, which consists of $50$ 2D-radar sensors and $50$ position sensors (such as cameras) in an area of $[-1000, 1000]~\SI{}{\meter}\times[-1000, 1000]~\SI{}{\meter}$, to track an unknown number of objects.
With this sensor network, two scenarios are constructed with $35$ and $50$ mobile objects, respectively. The ground truth and positions of sensors are shown in Figure~\ref{fig:s2_truth}.
In particular, all radar sensors have a maximum detection range of $\SI{150}{\meter}$ and detection probability of $P_{D}^{\text{(radar)}}=0.98$ and clutter rate $\lambda_{c}^{\text{(radar)}}=0.1$, while all position sensors have a maximum detection range of $\SI{300}{\meter}$, detection probability of $P_{D}^{\text{(pos)}} = 0.7$, and clutter rate $\lambda_{c}^{\text{(pos)}} = 1$. The total simulation duration is $75$ time steps.

\vspace{0.2cm}
\noindent\textbf{Object dynamic model.~}Each object follows a constant velocity (CV) motion model, given by
\begin{align} \label{eq:object_dynamic_model}
    x_k = F^{CV} x_{k-1} + q^{CV}_{k-1}.
\end{align}
where $x_k$ is the object state at time $k$ with $$x = [\mathrm{x},\dot{\mathrm{x}},\mathrm{y},\dot{\mathrm{y}}]^T,$$ comprising of its position $[\mathrm{x},\mathrm{y}]^T$ and velocity $[\dot{\mathrm{x}},\dot{\mathrm{y}}]^T$ in the 2D Cartesian coordinate system. Here, $F^{CV} = [1,T_0; 0, T_0] \otimes I_2 $ where $T_0=1$~s is the measurement sampling interval,  $\otimes$ denotes Kronecker tensor product, and $I_2$ denotes a $2\times 2$ identity matrix, and $q^{CV}$ is a $4\times 1$ zero mean Gaussian process noise,~\ie, $q^{CV} \sim \mathcal{N}(0_{4\times1}, \Sigma^{CV})$, with $\Sigma^{CV} = \sigma^2_{CV} [T^3_0/3,T^2_0/2;T^2_0/2,T_0] \otimes I_2$ and $\sigma_{CV} = 5$m/s\textsuperscript{2}.

\vspace{0.2cm}
\noindent\textbf{Measurement Model.~} 
The measurement functions for the 2D-radar sensor and 2D-position sensor are as follows:

\noindent\textit{i) 2D radar sensor.~}Each detected object $x$ yields a 2D radar measurement, consisting of range, range rate, and azimuth, \ie
\begin{align}
    z_k = h^{(\text{radar})}(x_k) + \eta^{(\text{radar})}_k,
\end{align}
where 
\begin{align}
h^{(\text{radar})}(x)= \begin{bmatrix}
     \sqrt{\mathrm{x}^2 + \mathrm{y}^2  } \\
     \big(\mathrm{x} \dot{\mathrm{x}} + \mathrm{y} \dot{\mathrm{y}}  \big) /\sqrt{\mathrm{x}^2 + \mathrm{y}^2 } \\
     \mathrm{atan2}\big( \mathrm{y}/\mathrm{x} \big)
    \end{bmatrix}
\end{align}
and $\eta^{(\text{radar})}_k$ is a $3\times1$ zero mean Gaussian measurement noise,~\ie, $\eta^{(\text{radar})}_k \sim \mathcal{N}(0_{3\times1},\sigma^{(\text{radar})}_\eta{\sigma^{(\text{radar})}_\eta}^{T})$, with $\sigma^{(\text{radar})}_\eta = $ $[10~\text{m},2~\text{m/s},\pi/180~\text{rad}]^T$.

\noindent\textit{ii) 2D position sensor.~} Each detected object $x$ yields a 2D position measurement, consisting of $x$-coordinate and $y$-coordinate, \ie
\begin{align}
    z_k = h^{(\text{pos})}(x_k) + \eta^{(\text{pos})}_k,
\end{align}
where 
\begin{align}
h^{(\text{pos})}(x)= [ \mathrm{x}, \mathrm{y}]^T \label{eq:pos_meas_func}
\end{align}
and $\eta^{(\text{pos})}_k$ is a $2\times1$ zero mean Gaussian measurement noise,~\ie, $\eta^{(\text{pos})}_k \sim \mathcal{N}(0_{2\times1},{\sigma^{2}_{\eta}}^{(\text{pos})} I_{2})$, with $\sigma^{(\text{pos})}_\eta = \SI{10}{\meter}$. 

\vspace{0.2cm}
\noindent\textbf{Experiments and Results.~}In the following, we employ the experimental scenario and settings in this section for several investigations described below.
\begin{enumerate}[label=\roman*)]
    \item Evaluate and compare the tracking performance of the proposed DMOT method.
    \item Compare the computational cost of the fusion methods.
    \item Demonstrate the scalability of the proposed approach with respect to an increasing number of sensors.
    \item Demonstrate the impact of parameter selection on tracking performance when employing DBSCAN or Mean-Shift clustering compared to the proposed CDP-based analysis.
\end{enumerate}

\begin{figure}[!tb]
\centering \includegraphics[width=1.0\textwidth]{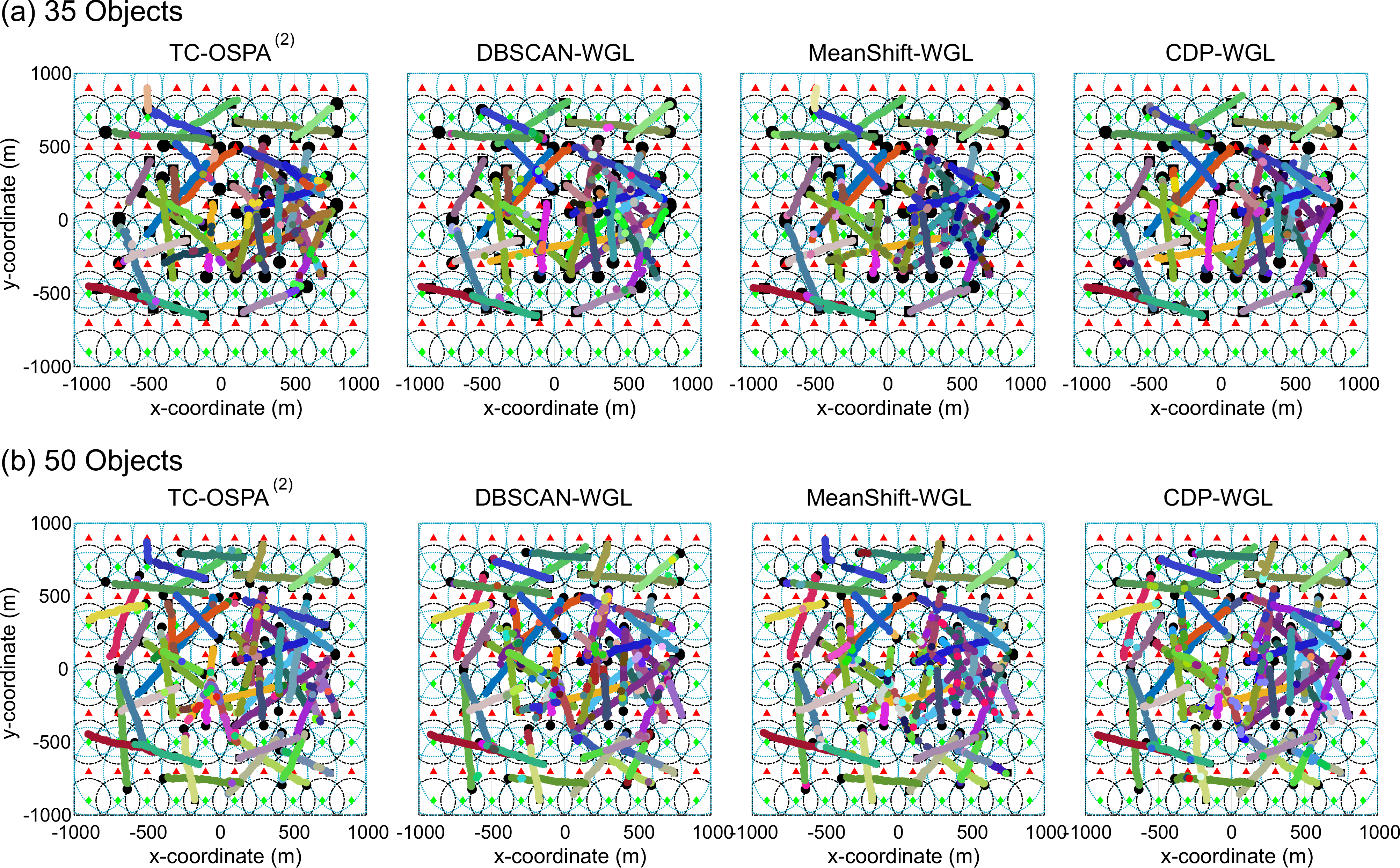}
\vspace{-1cm}
\caption{Heterogeneous sensors scenario: an instance of estimated trajectories using TC-OSPA\textsuperscript{(2)}, \dbscan{}, \meanshift{} and \cdp{} algorithms with LMB filter when tracking objects at each sensor node (a) with $35$ mobile objects; (b) with $50$ mobile objects.}
\label{fig:s2_est_sample} 
\end{figure}

\noindent\textit{Tracking Performance.~}
Figure~\ref{fig:s2_est_sample} depicts the labelled multi-object state estimates at node $1$ for TC-OSPA\textsuperscript{(2)}, \dbscan{}, \meanshift{} and \cdp{} fusion methods, respectively, obtained from one MC simulation execution.
Visually, all three methods are able to successfully estimate the positions of all objects but with varying degrees of label consistency as discussed in the overall results below.

\begin{figure}[!tb]
\centering \includegraphics[width=1.0\textwidth]{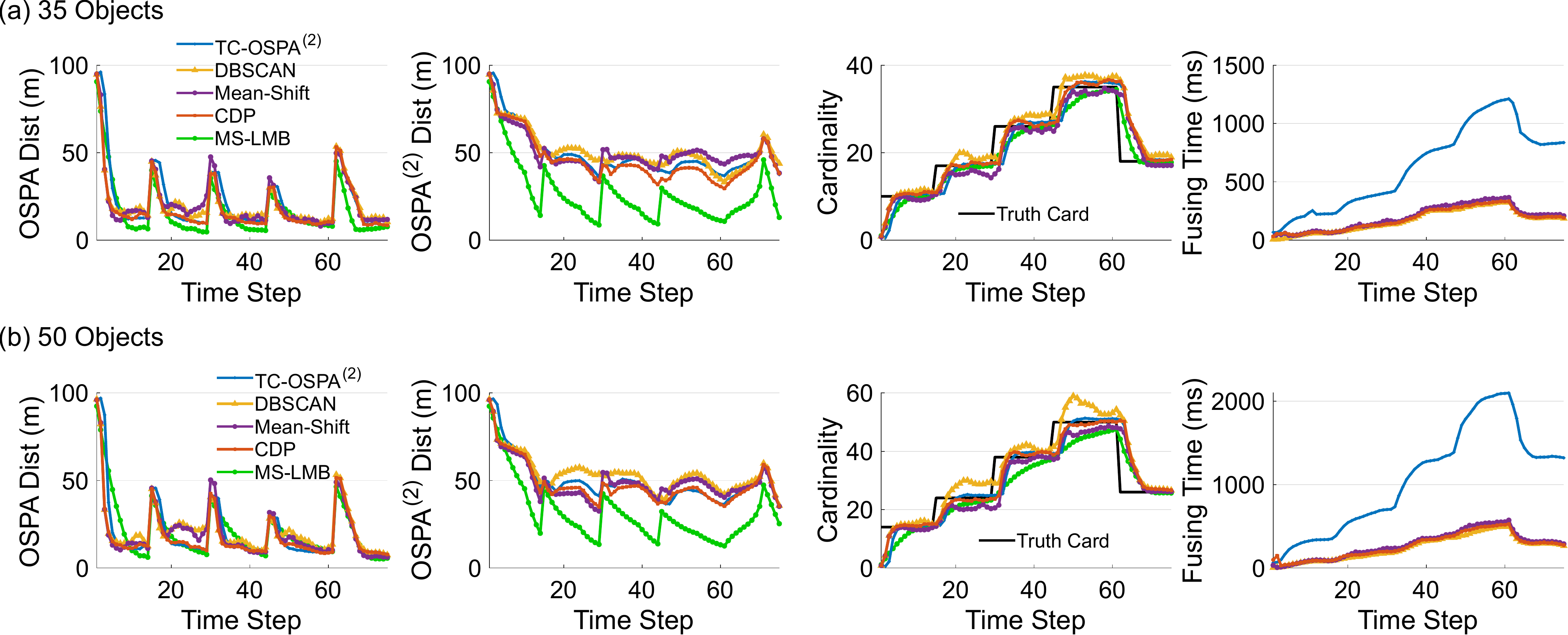}
\vspace{-0.8cm}
\caption{Heterogeneous sensors scenario: average OSPA, OSPA\textsuperscript{(2)} Distance, Cardinality and fusing time over $100$ MC runs using LMB filter when tracking (a) $35$ mobile objects; (b) $50$ mobile objects.}
\vspace{-0.3cm}
\label{fig:s2_result} 
\end{figure}

\begin{table}[!tb]
\centering
\small
\caption{Heterogeneous sensors scenario. The results reported are obtained from over 100 Monte Carlo executions for each method. Bold text indicates the best result.}
\vspace{2mm}
\label{tab:s2_table}
\begin{tabular}{lllll}
\toprule
\multicolumn{5}{c}{\textbf{Heterogeneous sensors: 35 Objects}}\\
\hline
Method & OSPA (m) & OSPA\textsuperscript{(2)} Dist (m) & Fusing time (ms) & Speed up\\ 
\hline
\hline
TC-OSPA\textsuperscript{(2)}  & 21.81     & 49.50    & 626.35  & $\times$1    \\
DBSCAN-Unweighted  & 20.85 &   53.14     & 185.84  & $\times$3.37\\  
\dbscan{}              & 20.86       & 50.98    & \textbf{169.64}   &  $\times$3.69 \\
MeanShift-Unweighted  &   20.80 &  52.02 &   196.70 & $\times3.18$  \\
\meanshift{}  &  20.78 & 50.23 &   201.08&  $\times3.11$\\
CDP-Unweighted  & 18.96   & 50.59     &  185.10    & $\times$3.38\\
\cdp{}    & \textbf{18.86}       & \textbf{46.03}    & 179.15    &  $\times$3.49  \\
\hline
MS-LMB (Centralized \textit{baseline})   & {16.46 }   & {26.47}   &{NA}   &{NA}   \\
\midrule
\multicolumn{5}{c}{\textbf{Heterogeneous sensors: 50 Objects}}\\
\hline
TC-OSPA\textsuperscript{(2)}  & 20.66             & 49.76             & 1060.42  & $\times$1    \\
DBSCAN-Unweighted  & 21.47 &   54.51       & 255.15 & $\times$4.16\\   
\dbscan{} & 21.46    & 53.63      & \textbf{245.03}   &  $\times$4.33        \\
MeanShift-Unweighted  &   20.32 &  50.90 &   288.25 & $\times3.68$ \\
\meanshift{}  & 20.19 &  49.18 &  279.62 & $\times3.79$  \\
CDP-Unweighted  & 18.69   & 50.57    &  267.74    & $\times$3.96\\
\cdp{}    & \textbf{18.65}       & \textbf{47.95}    & 263.64   &  $\times$4.02   \\
\hline
MS-LMB (Centralized \textit{baseline})    & {20.85}            & {31.30}             &{NA}                 &{NA}                      \\
\bottomrule
\end{tabular}
\end{table}

Figure~\ref{fig:s2_result} and Table~\ref{tab:s2_table} present detailed tracking performance comparisons across all MC executions.
As seen in Figure~\ref{fig:s2_result}, all clustering-based methods have better estimation accuracy than the TC-OSPA\textsuperscript{(2)} method at time points where the cardinality increases (for example, at approximately $\SI{15}{\second}, \SI{30}{\second}$ and $\SI{50}{\second}$ as shown in the cardinality plot) indicated by their lower OSPA distance measures; but performance improves to be similar to the proposed \cdp{} method at later points in time. Clustering approaches are observed to be more effective at these time instances because the TC-OSPA\textsuperscript{(2)} method can only fuse local tracks with a minimum length of $2$, which reduces responsiveness during object births. Consequently, in the summary results in Table~\ref{tab:s2_table}, cluster analysis methods are seen to perform better than TC-OSPA\textsuperscript{(2)} in the OSPA metric.

Then, as seen in Figure~\ref{fig:s2_result}, \cdp{} is seen to have more accurate cardinality estimates while \dbscan{} overestimates and \meanshift{} underestimates the number of objects. These results demonstrate the advantages of the CDP-based analysis algorithm. Even though the cardinality estimates for TC-OSPA\textsuperscript{(2)} are observed to lag those of our proposed \cdp{} method as seen Figure~\ref{fig:s2_result}, due to the delay in object confirmations in TC-OSPA\textsuperscript{(2)} method, over time these estimates become accurate.  

In terms of tracking performance, the overall results in Table~\ref{tab:s2_table} as well as the detailed results in Figure~\ref{fig:s2_result} show our \cdp{} to attain the lowest OSPA\textsuperscript{(2)} distance value. This result can be attributed to the higher estimation accuracy of our CDP-based kinematic consensus algorithm and lower track label switching (label consistency) of the weighted graph label consensus algorithm. Importantly, a noticeable decrease in OSPA\textsuperscript{(2)} can be observed when comparing each clustering-based method with their respective counterparts using the unweighted label consensus algorithm---\undbscan{}, \unmeanshift{} and \uncdp{}. This result demonstrates the lower track label switching (or higher label consistency) achieved with our proposed weighted graph label consensus algorithm.

\vspace{3mm}
\noindent\textit{Computational Costs.~}
A key motivation for our formulation is to address the computational complexity posed by the difficult and complex labelled density fusion problem faced in DMOT. The fusing time (and speed up) results in Figure~\ref{fig:s2_result} and Table~\ref{tab:s2_table} demonstrate TC-OSPA\textsuperscript{(2)} method to demand substantial time for computing fusion results---globally consistent tracks. In contrast, the clustering-based methods use, on average, only $27\%$ of the fusing time required by TC-OSPA\textsuperscript{(2)}. This highlights the efficiency gains of our proposed clustering-based approaches in mitigating the complexity of the multi-dimensional assignment problem we discussed in Section~\ref{sec:intro} whilst achieving similar or better tracking performance compared to TC-OSPA\textsuperscript{(2)} as summarised in Table~\ref{tab:s2_table}.

\begin{figure}[!tb]
\centering \includegraphics[width=1.0\textwidth]{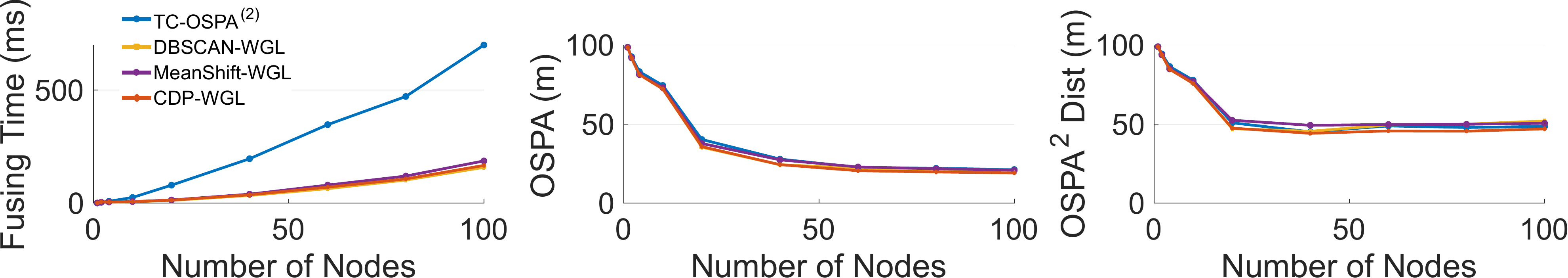}
\vspace{-0.7cm}
\caption{Heterogeneous sensors scenario (35 objects): tracking performance at one sensor node over 100 MC runs when the number of connected sensor nodes increases from $1$ to $100$.}
\label{fig:s2_varSensor} 
\end{figure}

\vspace{3mm}
\noindent\textit{Scalability.~}
An important facet of a distributed algorithm is the ability to scale to larger problem dimensions without significant increases in computational costs. Figure~\ref{fig:s2_varSensor} depicts the overall tracking performance of TC-OSPA\textsuperscript{(2)}, \dbscan{}, \meanshift{} and \cdp{} at one sensor node as the number of interconnected sensor nodes is increased from $3$ to $100$.
The results validate the scalability of our proposed fusion strategy, wherein the fusion time increases linearly with respect to the number of connected sensor nodes.
Additionally, the clustering methods require lower fusing times than TC-OSPA\textsuperscript{(2)} in all settings due to their lower computational complexity.
Notably, as expected, when the number of nodes increase, the OSPA and OSPA\textsuperscript{(2)} errors decrease since the nodes can benefit from shared local labelled multi-object state estimates of other nodes to complement their own limited FoVs, thereby improving coverage area and tracking accuracy.

\begin{figure}[!tb]
\centering \includegraphics[width=0.8\textwidth]{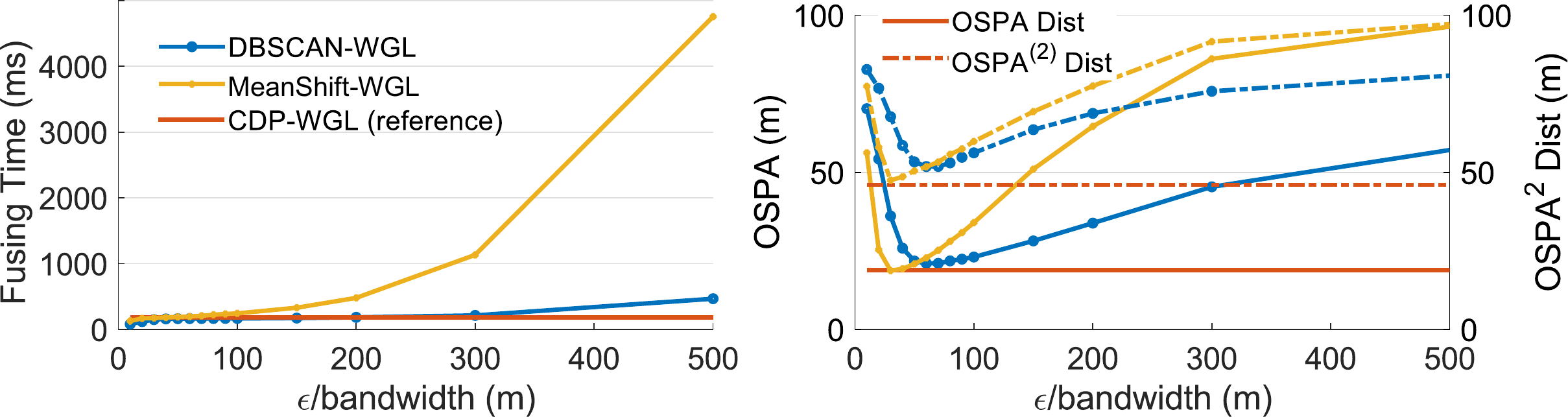}
\vspace{-0.3cm}
 \caption{Heterogeneous sensors scenario comparing parameter selection sensitivity of \dbscan{} and \meanshift{} on tracking performance and fusing time with parameter-less CDP-WGL method. Results are at one sensor node over $100$ MC runs when the \dbscan{} method's parameter $\epsilon$ and \meanshift{} method's bandwidth are increased.}
\label{fig:s2_dbscan_sensitivity} 
\end{figure}

\vspace{3mm}
\noindent\textit{Hyper Parameter Sensitivity.~}
Figure~\ref{fig:s2_dbscan_sensitivity} shows the tracking performance at one sensor node over $100$ MC runs as the $\epsilon$ parameter for \dbscan{} and the bandwidth for \meanshift{} is changed from $\SI{2.5}{\meter}$ to $\SI{500}{\meter}$, compared against our proposed \cdp{} method.
The \dbscan{} and \meanshift{}'s tracking performance is sensitive to the selection of $\epsilon$ and bandwidth, respectively. Consequently, inappropriately chosen parameters can significantly degrade the performance of these methods.
By comparison, our proposed \cdp{} method is not encumbered by parameter selection and is able to outperform \dbscan{} and \meanshift{} methods even under optimal value settings for $\epsilon$/bandwidth for the two methods. The results further highlight the advantages of the \cdp{} method in terms of robustness  and ease of deployment.
In terms of fusing time, the \dbscan{} method is only slightly faster than the \cdp{} in the best-case scenario. Considering the respective computation complexity of the two clustering algorithms as shown in Table~\ref{tab:algo_complexity}, the difference is most likely due to implementation differences and/or the lower average time complexity of the DBSCAN algorithm.
Comparatively, the \meanshift{} algorithm suffers the most from incorrect bandwidth selection. When a large bandwidth is used, the \meanshift{} algorithm groups more local estimates into the same cluster, which increases the number of network edges in the weighted label graph, \ie{} generating increased numbers of label associations. Consequently, increasing the amount of time spent on updating the graph relative to the other clustering algorithms.

\begin{figure}[!tb]
\centering \includegraphics[width=0.8\textwidth]{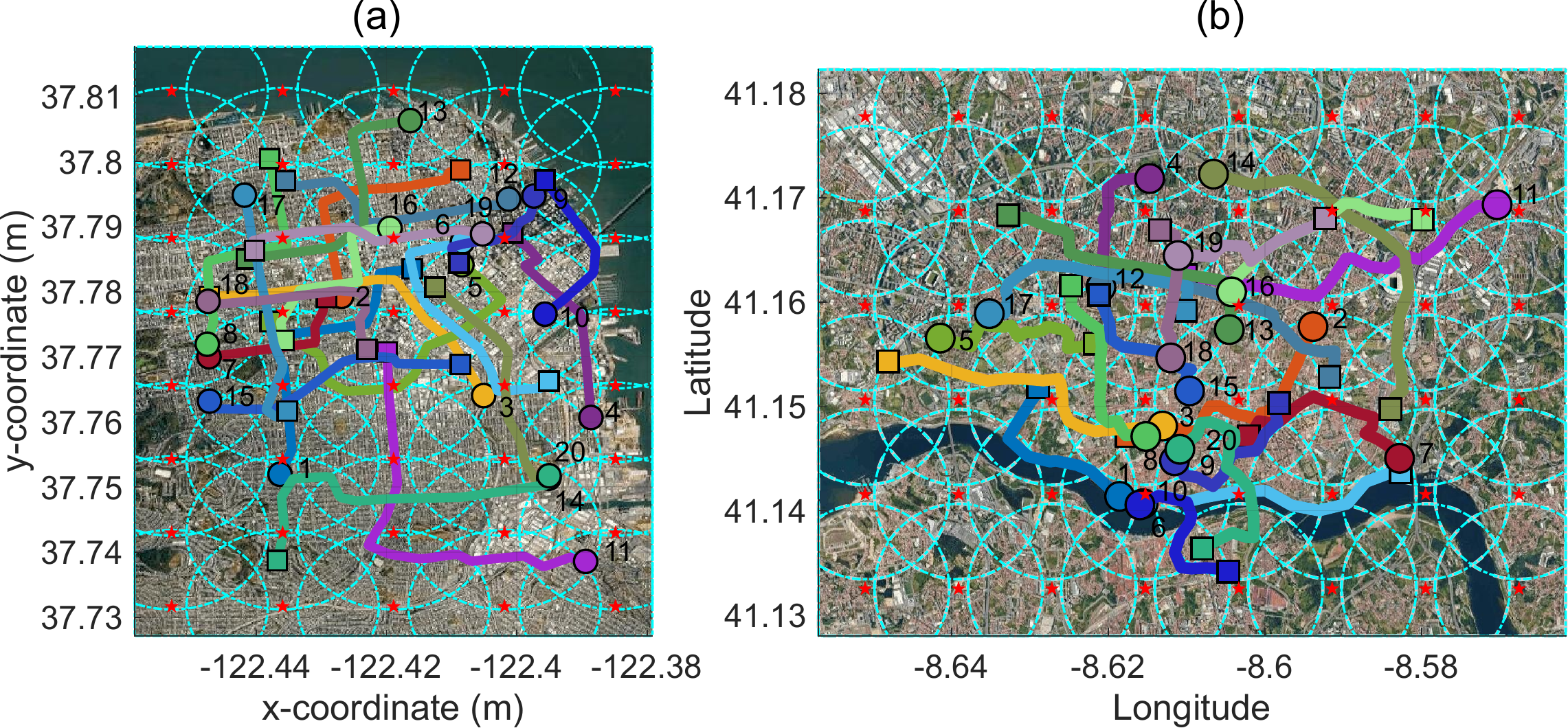}
\vspace{-0.3cm}
\caption{(a) San Francisco taxi tracking scenario:  the ground truth trajectories of $20$ Taxis tracked by $40$ sensor nodes.  (b) Porto taxi scenario: the ground truth trajectories of $20$ Taxis tracked by $48$ sensor nodes. Starting and stopping positions are denoted by $\ocircle$ and $\Box$.}
\vspace{-0.3cm}
\label{fig:taxi_truth} 
\end{figure}

\subsection{Taxi Tracking}\label{sec:taxi-trad-exp}
In this scenario, we evaluate the performance of our proposed distributed algorithm under real-world trajectory settings using the CRAWDAD taxi dataset~\cite{epfl-mobility-20090224} and the Porto taxi dataset~\cite{pkdd-15}. The CRAWDAD taxi data set contains taxi trajectories in the San Francisco Bay Area, USA and the Porto taxi dataset consists of taxi trajectories in Porto, Portugal, released in the ECML/PKDD 2015 challenge competition. We selected $20$ taxi tracks from each dataset and followed a similar approach to~\cite{Hoa2024TSPPlanForTrack} to speed up the time duration of the ground truth by a factor of $10$.

The simulated environment using the CRAWDAD dataset has an area of $\SI{7000}{\meter}\times\SI{10000}{\meter}$ and uses a total tracking period of $\SI{120}{\second}$ with $40$ limited FoV sensor nodes placed uniformly in the area with a maximum detection range of $\SI{1300}{\meter}$, a detection probability $P_{D} = 0.9$ and a clutter rate $\lambda_{c} = 10$.
The simulated environment using the Porto taxi dataset has an area of $\SI{8000}{\meter}\times\SI{6000}{\meter}$ and uses a total tracking period time of $\SI{115}{\second}$ with $48$ limited FoV sensor nodes placed uniformly in the area with a maximum detection range of $\SI{900}{\meter}$, a detection probability $P_{D} = 0.9$ and a clutter rate $\lambda_{c} = 10$. The ground truth tracks and the positions of sensor nodes of both scenarios are shown in Figure~\ref{fig:taxi_truth}. All sensor nodes are assumed to be network-connected in both scenarios.

\vspace{0.2cm}
\noindent\textbf{Object Dynamic Model.~}To approximate the behaviour of taxis', we employ the constant turn (CT) model with an unknown turn rate as the dynamic model. This model is defined as follows: 
\begin{align} \label{eq:ct_model}
\begin{split}
    x_{k} &= F^{CT}(\theta_{k})x_{k-1} + G^{CT}\omega_{k} \\
    \theta_{k} &= \theta_{k-1} + T_{0}q_{k}, \\
\end{split}
\end{align}
where $x_{k}$ is the object state at time $k$ with $x = [\mathrm{x}, \dot{\mathrm{x}}, \mathrm{y}, \dot{\mathrm{y}}, \theta]$, consisting of its 2D position, velocity and turning rate, $T_{0} = \SI{1}{\second}$, $\omega_{k}\sim \mathcal{N}(0_{4\times1}, \sigma_{\omega}^{2}I_{2})$ with $\sigma_{\omega} = \SI{5}{\meter/\second^{2}}$, $q_{k}\sim\mathcal{N}(0, \sigma_{q}^{2})$, and $\sigma_{q} = (\pi/60)\SI{}{\radian/\second^{2}}$,
\begin{align}
        F^{CT}(\theta) = \begin{bmatrix}
                         1  &  \frac{\sin(\theta T_{0})}{\theta} & 0 & -\frac{1 - \cos(\theta T_{0})}{\theta} \\
                         0  &  \cos(\theta T_{0})  &  0  & -\sin(\theta T_{0})  \\
                         0  & \frac{1-\cos(\theta T_{0})}{\theta}  & 1  &  \frac{\sin(\theta T_{0})}{\theta}  \\
                         0  & \sin(\theta T_{0})  &  0  &  \cos(\theta T_{0}) \\
                         \end{bmatrix},\quad 
        G^{CT} = \begin{bmatrix}
                 \frac{T_{0}^{2}}{2}  & 0 \\
                 T_{0}  & 0 \\
                 0  &  \frac{T_{0}^{2}}{2} \\
                 0  &  T_{0}^{2} \\
                 \end{bmatrix}.
\end{align}     
\noindent\textbf{Measurement Model.~}Each local node comprises a 2D-position sensor with its measurement function specified in \eqref{eq:pos_meas_func} with measurement noise $\sigma^{(\text{pos})}_\eta = \SI{10}{\meter}$.

\noindent\textbf{Experiments and Results.~}In the following, we employ the real-world trajectory dataset for two investigations described below.
\begin{itemize}
\item[i)] Evaluate and compare the tracking performance of the proposed DMOT method and demonstrate the generalisation of the method.
\item[ii)] Investigate the effectiveness of the proposed global label consensus algorithm based on the WGL method.
\end{itemize}

\begin{figure}[!tb]
\centering \includegraphics[width=1.0\textwidth]{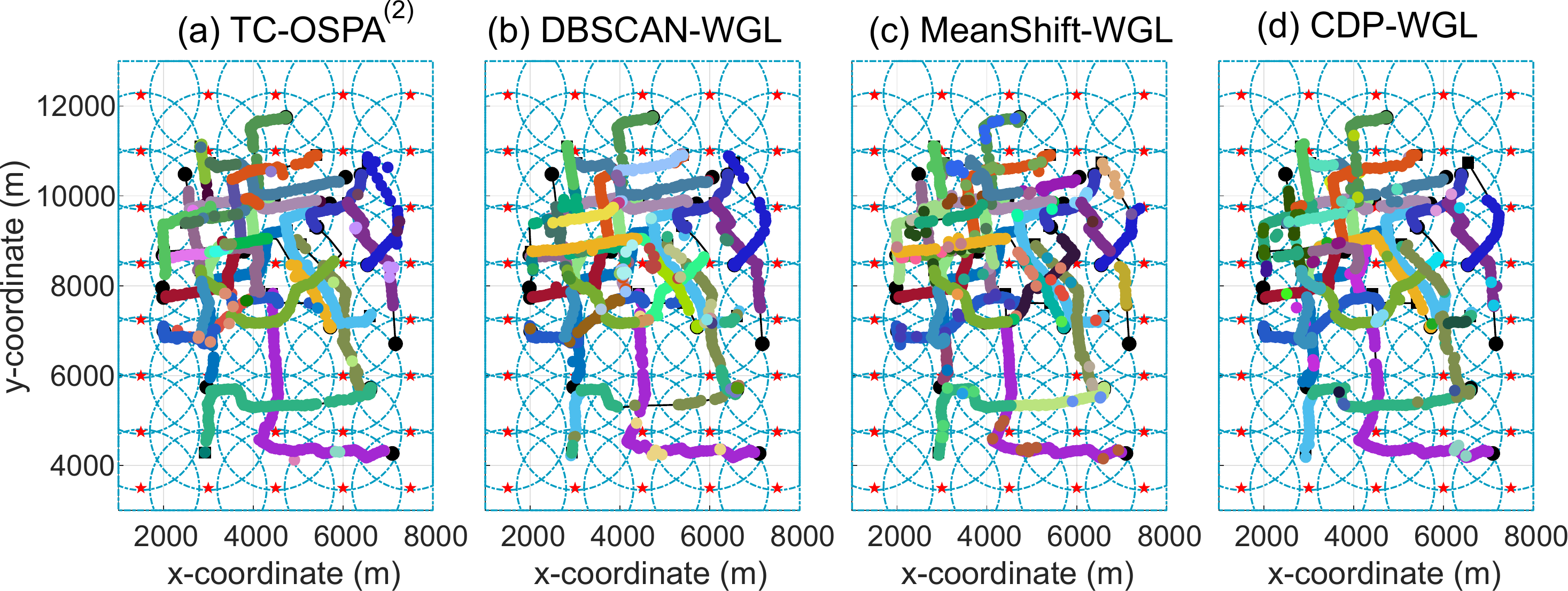}
\vspace{-0.8cm}
\caption{San Francisco taxi tracking scenario: an instance of estimated trajectories using TC-OSPA\textsuperscript{(2)}, \dbscan{}, \meanshift{} and \cdp{} algorithms with LMB filter.}
\vspace{-0.4cm}
\label{fig:taxi_est_sample} 
\end{figure}

\begin{figure}[!tb]
\centering \includegraphics[width=1.0\textwidth]{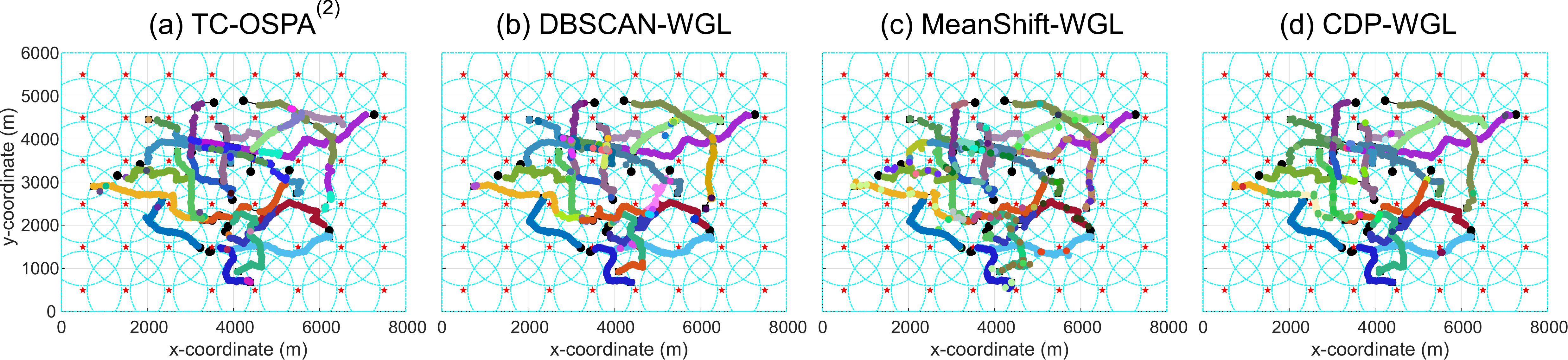}
\vspace{-0.8cm}
\caption{Porto taxi tracking scenario: an instance of estimated trajectories using TC-OSPA\textsuperscript{(2)}, \dbscan{}, \meanshift{} and \cdp{} algorithms with LMB filter.}
\vspace{-0.4cm}
\label{fig:ECML_est_sample} 
\end{figure}

\noindent\textit{Tracking Performance (Results Generalisation).~}
For illustration, Figure~\ref{fig:taxi_est_sample} and \ref{fig:ECML_est_sample} depict the labelled multi-object state estimates at node $1$ for TC-OSPA\textsuperscript{(2)}, \dbscan{}, \meanshift{} and \cdp{} fusion methods, respectively, for a single tracking experiment result. All three methods successfully estimate the taxi's positions with varying degrees of label consistency discussed further in the detailed results.

\begin{figure}[!tb]
\centering \includegraphics[width=1.0\textwidth]{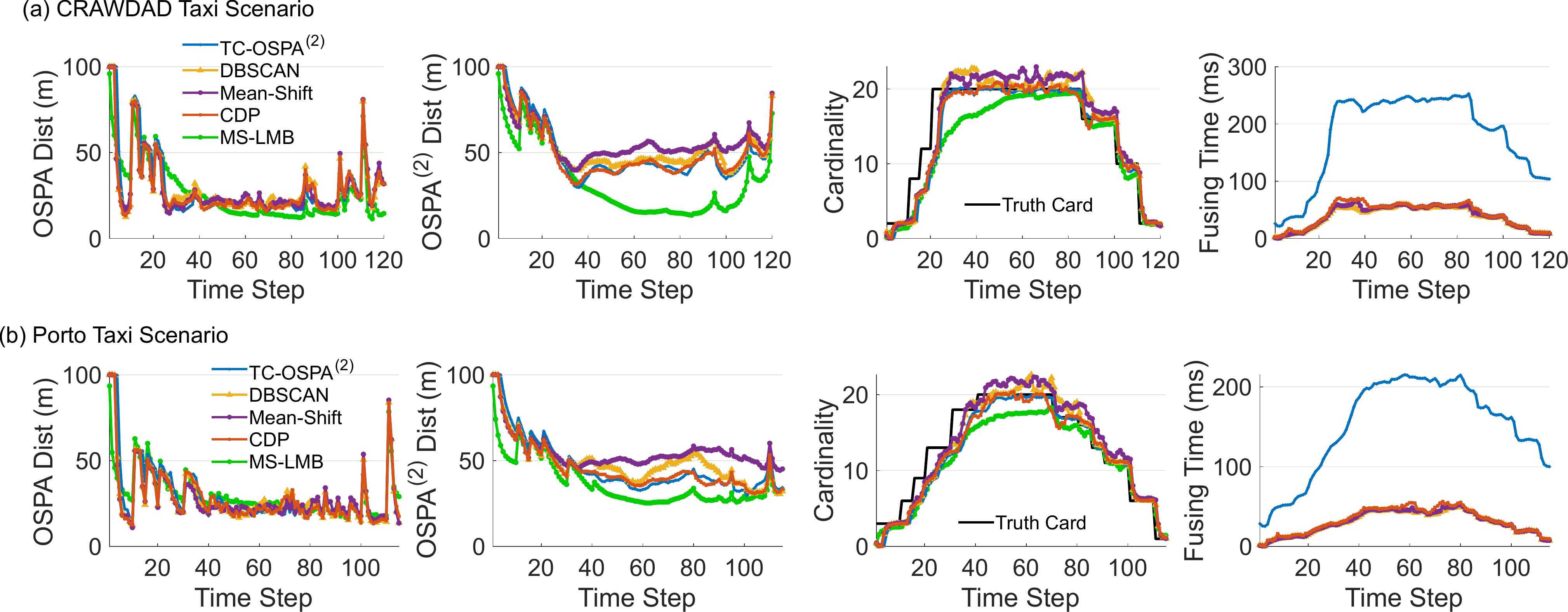}
\vspace{-0.8cm}
\caption{Averaged OSPA, OSPA\textsuperscript{(2)} Distance, Cardinality and fusing time over $100$ MC runs where each node employs an LMB filter for (a) San Francisco taxi tracking scenario;  (b) Porto taxi tracking scenario.}
\label{fig:taxi_stat} 
\end{figure}

\begin{table}[!tb]
\centering
\small
\caption{San Francisco and Porto taxi tracking scenario result. The results reported are obtained from 100 Monte Carlo trials for each method. Bold text indicates the best result.}
\vspace{2mm}
\label{tab:taxi_table}
\begin{tabular}{lllll}
\toprule
\multicolumn{5}{c}{\textbf{San Francisco Taxi Tracking Scenario}}\\
\hline
Method & OSPA (m) & OSPA\textsuperscript{(2)} Dist (m) & Fusing time (ms) & Speed up\\ 
\hline
\hline
TC-OSPA\textsuperscript{(2)}   & 29.72             & 50.28           & 180.13  & $\times$1     \\
DBSCAN-Unweighted  & 29.12     & 55.74      & 51.64  & $\times$3.49\\ 
\dbscan{}                   & 29.29     & 52.53      & \textbf{37.59}  & $\times$4.79\\
MeanShift-Unweighted  &   29.21 &  60.43 &   41.45 & $\times4.35$  \\
\meanshift{}  &   29.28 &  56.61 &  39.53 & $\times4.56$  \\
CDP-Unweighted  & 28.71   & 53.64     &  49.18    & $\times$3.66\\ 
\cdp{}    & \textbf{28.66}     & \textbf{49.88}    & 43.65          & $\times$4.13\\
\hline
MS-LMB (Centralized \textit{baseline})  & {22.90}    & {29.77}    & {NA}    & {NA}\\
\midrule
\multicolumn{5}{c}{\textbf{Porto Taxi Tracking Scenario}}\\
\hline
TC-OSPA\textsuperscript{(2)}  & 29.80             & 46.19             & 150.78  & $\times$1    \\
DBSCAN-Unweighted  & 28.38 &   50.69       & 31.12 & $\times$4.85\\   
\dbscan{} & 28.29                & 48.62      & \textbf{29.92}   &  $\times$5.04        \\
MeanShift-Unweighted  &   28.43 &  56.47 &   32.13 & $\times4.69$ \\
\meanshift{}  & 28.41 &  54.67 &  30.70 & $\times4.91$  \\
CDP-Unweighted  & \textbf{28.19}   & 47.63    &  33.22    & $\times$4.54\\
\cdp{}    & 28.27       & \textbf{45.32}    & 32.73            &  $\times$4.61          \\
\hline
MS-LMB (Centralized \textit{baseline})    & {29.94}            & {36.76}             &{NA}                 &{NA}                      \\
\bottomrule
\end{tabular}
\end{table}

\begin{figure}[!tb]
\centering \includegraphics[width=0.35\textwidth]{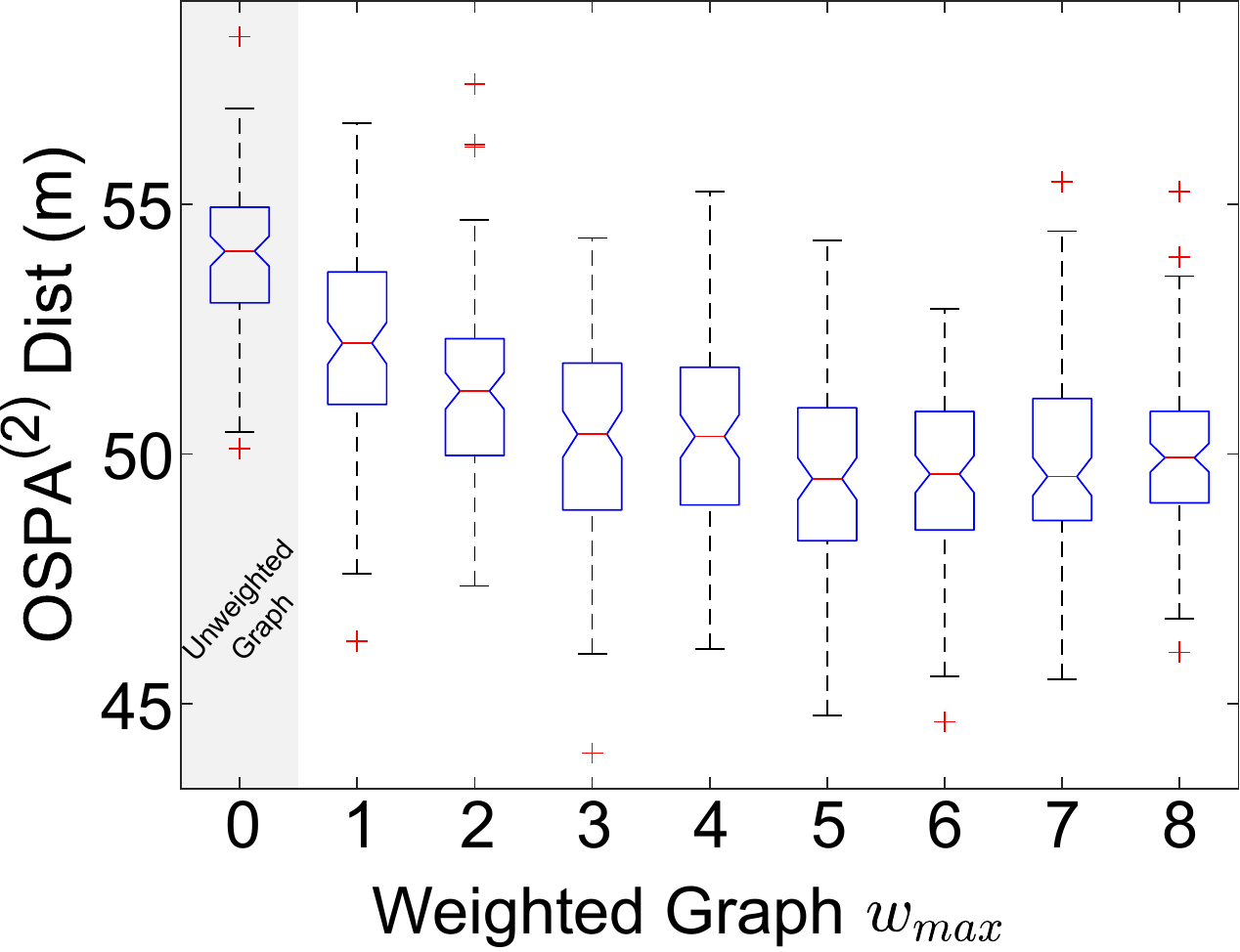}
\vspace{-0.3cm}
\caption{San Francisco taxi tracking scenario: OSPA\textsuperscript{(2)} Distance results with our proposed \cdp{} method employing different graph weights $w_{max}$ obtained from $100$ MC runs. The \textit{Unweighted Graph} ($w_{max} = 0$) corresponds to the label consensus method in~\cite{hoa2021distributed}.}
\vspace{-0.2cm}
\label{fig:weighted_graph_stat} 
\end{figure}

Figure~\ref{fig:taxi_stat} and Table~\ref{tab:taxi_table} present detailed performance comparisons across all MC executions of both taxi tracking scenarios.
Overall, despite the challenging dynamic behaviour of taxis, our \cdp{} algorithms perform better than TC-OSPA\textsuperscript{(2)} and the alternative cluster analysis methods \dbscan{} and \meanshift{} methods in terms of estimation and tracking accuracy as shown in the OSPA and OSPA\textsuperscript{(2)} results, respectively, in Table~\ref{tab:taxi_table}. In sympathy with results observed in Section~\ref{sec:hetero-exp}, the combination of cluster analysis with WGL can improve tracking performance---achieve a lower OSPA\textsuperscript{(2)}---through better global label consistency. These results confirm the robustness of the proposed WGL method and our DMOT algorithm-\cdp{}. Importantly, the results confirm the capacity to rely on the CPD-based clustering analysis approach, without the need for hyper-parameter determination, to more consistently resolve the multi-dimensional assignment problem across different scenarios.

As observed before, in Section~\ref{sec:hetero-exp}, \cdp{}'s cardinality estimates shown in Figure~\ref{fig:taxi_stat} are significantly more stable and accurate than the \dbscan{} and \meanshift{} methods, while \dbscan{} and \meanshift{} based analysis leads to overestimation of the cardinality.
Additionally, based on the fusing time results, we can observe the clustering-based methods to consistently require less than $28\%$ of the fusing time compared to TC-OSPA\textsuperscript{(2)}, confirming the expected efficiency gains of our proposed method.

\vspace{3mm}
\noindent\textit{Effectiveness of the Proposed Label Consensus Method.}
Investigation in Section~\ref{sec:hetero-exp} and Section~\ref{sec:taxi-trad-exp} demonstrated the impact of the WGL method to improve global label consistency across a distributed network of sensor nodes. Figure~\ref{fig:weighted_graph_stat} depicts the tracking performance at one sensor node in the San Francisco taxi tracking scenario as the parameter $w_{max}$ in the weighted graph label consensus method is increased from $0$ (equivalent to the unweighted graph label consensus method in \cite{hoa2021distributed}) to $8$. Consistent with our previous results, the weighted graph-based label consensus algorithm is observed to have significantly lower OSPA\textsuperscript{(2)} values for graph weight $w_{max} > 0$---i.e. compared to the unweighted graph method. Further, performance is robust to variation in $w_{max} \geq 3$. This demonstrates the effectiveness of the weighted graph method in reducing label inconsistency and the resulting improvements in OSPA\textsuperscript{(2)} values.

\vspace{3mm}
\noindent\textit{Summary.} 
Overall, the results further demonstrate the performance advantage of \cdp{} in a realistic DMOT setting. In this scenario, where the object dynamic model does not completely match the motion of the underlying objects, \cdp{} is demonstrated to provide better estimation and tracking accuracy whilst performing $4.13$ to $4.61$ times faster than the TC-OSPA\textsuperscript{(2)} method.

\section{Conclusions} \label{sec:conclusion}
In this paper, we presented a clustering-based approach for DMOT problems under limited and unknown FoV sensors. Our solution comprises a clustering-based kinematic fusion method with a weighted graph-based label-consistent fusion method.
Detailed comparative analyses using varied scenarios demonstrated that our clustering solution not only reduces computational demands but also maintains competitive tracking accuracy compared to state-of-the-art track consensus algorithms. Furthermore, the adaptability of our methods was particularly evident in realistic DMOT settings, even when faced with dynamic model discrepancies. This work demonstrates the potential benefits and effectiveness of embedding clustering algorithms into distributed multi-object tracking systems, especially in environments with restricted sensor visibility. Notably, we can observe the challenging problem of determining hyperparameters, such as $\epsilon$ in DBSCAN, in the context of limited FoV sensors with partially overlapping FoV, still remains. Our work suggests that determining such a hyperparameter in an online manner forms a promising direction for future work.

\section*{Acknowledgement}
This study was partially supported by the Defence Science and Technology Group (DSTG), Australia.

\bibliographystyle{elsarticle-num}
\bibliography{references}

\end{document}